\documentclass[11pt, a4paper]{article}
\usepackage[a4paper, total={6.5in, 8.5in}]{geometry}
\usepackage{amsmath,amsfonts}
\usepackage{algorithmic}
\usepackage{algorithm}
\usepackage{array}
\usepackage{mathtools, cuted}
\usepackage{subfigure}
\usepackage{textcomp}
\usepackage{stfloats}
\usepackage{url}
\usepackage{verbatim}
\usepackage{graphicx}
\usepackage{cite}
\usepackage{authblk}
\usepackage{stfloats}

\begin{document}

\title{ESPiM: Eye-Strain Probation Model, An Eye-Tracking Analysis Measure for Digital Displays}

\author{Mohsen Parisay, Negar Haghbin\textsuperscript{a}, Charalambos Poullis\textsuperscript{a,b}, and Marta Kersten-Oertel\textsuperscript{a,c}
}
\affil{\textsuperscript{a}Department of Computer Science and Software Engineering (m\_parisa@encs.concordia.ca, n\_haghbi@encs.concordia.ca);
\textsuperscript{b}Immersive and Creative Technologies Lab (charalambos@poullis.org);
\textsuperscript{c}PERFORM Center (marta@ap-lab.ca); Concordia University, Montreal, Canada}


\markboth{IEEE Transactions on Human-Machine Systems}%
{Shell \MakeLowercase{\textit{et al.}}: A Sample Article Using IEEEtran.cls for IEEE Journals}

\date{}
\maketitle

\begin{abstract}
Eye-strain is a common issue among computer users due to the prolonged periods they spend working in front of digital displays. This can lead to vision problems, such as irritation and tiredness of the eyes and headaches. We propose the Eye-Strain Probation Model (ESPiM), a computational model based on eye-tracking data that measures eye-strain on digital displays based on the spatial properties of the user interface and display area for a required period of time. As well as measuring eye-strain, ESPiM can be applied to compare (a) different user interface designs, (b) different display devices, and (c) different interaction techniques. Two user studies were conducted to evaluate the effectiveness of ESPiM. The first was conducted in the form of an in-person study with an infrared eye-tracking sensor with 32 participants. The second was conducted in the form of an online study with a video-based eye-tracking technique via webcams on users' computers with 13 participants. Our analysis showed significantly different eye-strain patterns based on the video gameplay frequency of participants. Further, we found distinctive patterns among users on a regular 9-to-5 routine versus those with more flexible work hours in terms of (a) error rates and (b) reported eye-strain symptoms.
\end{abstract}

\subsubsection*{Keywords}
\textit{Eye-strain, eye-tracking analysis, gaze-based interaction, Fitts' Law}    

\section{Introduction}

Today, in many parts of the world, many employees work sitting in front of computers for eight hours a day. The typical work atmosphere, which includes working in enclosed spaces, e.g. cubicles or offices, and dealing with multiple devices (e.g. PC, smartphone, telephone) simultaneously (i.e. multitasking), can lead to an increased workload. Yet, there is a strong correlation between excessive screen time and health risks such as cardiovascular diseases, impaired vision, and bone density \cite{screen_time_risks} and a correlation between long screen times of elementary school students with dry eyes (the malfunctioning of tear production in the eyes \cite{dry_eyes_definition}) and learning abilities \cite{screen_time_children}. 

Computer vision syndrome (CVS) is composed of multiple eye vision problems due to the prolonged use of digital displays, including tablets and smartphones. Other symptoms of CVS include \textit{eye-strain}, \textit{eye burn}, \textit{double vision}, and \textit{blurred vision} \cite{mowatt2018computer}. These CVS symptoms not only affect visual comfort but also are a major cause of work-related stress \cite{munshi2017computer} and productivity in both adults and teenagers \cite{rosenfield2016computer}. In this paper, we address eye-strain, also known as visual fatigue, as one of the significant CVS symptoms \cite{mowatt2018computer}. Eye-strain can be caused by moving images \cite{iwa2005image} and occurs when focusing on near objects. It is a common issue among computer users due to the prolonged periods they spend working in front of a monitor \cite{vasiljevas2016modelling}. Eye-strain can cause motion sickness \cite{kuze2008subjective}, vision problems such as irritation and tiredness of the eyes, and headaches \cite{ukai2008visual}. 

In this paper, we introduce the Eye-Strain Probation Model (ESPiM), an integrated measurement model for eye-strain based on target selection tasks relying on spatial targets' and screen properties via Fitts' Law and eye-tracking fixation points.

To demonstrate the efficacy of ESPiM, we measured target selection performance with data from two eye-tracking user studies, one in-person laboratory study and one online web application study. We considered the distinctive patterns among participants based on video gameplay frequency. Moreover, we studied eye-strain patterns, including eye symptoms among typical 9-to-5 participants and flexible (anytime beyond 9-to-5) participants and found the flexible groups experienced a higher number of eye symptoms than the 9-to-5 group. Moreover, we recorded significantly higher error rates for the 9-to-5 groups than the flexible group.

\section{Related Work}
In the following section, we describe Fitts' Law which is related to task difficulty of target selection tasks and used within the ESPiM model, as well as previous work in eye-strain and eye-tracking models.  

\subsection{Fitts' Law}
Originally proposed to measure task difficulty, Fitts' Law predicts the amount of movement time (MT) to activate a target based on a specific size (W) at a certain distance (D). Fitts' Law is formulated as: $MT = a + b\log_{2}(\frac{2D}{W})$, where \textit{a} and \textit{b} are empirically defined constant values. However, the Shannon formulation is most commonly used to calculate the index of difficulty in the field of HCI, $ID = \log_{2}(1 + \frac{D}{W})$ \cite{mackenzie1989note}. Among the earliest applications in HCI, Fitts' Law was used to compare input devices (mouse, joystick, step keys and text keys) for text selection on a display \cite{card1978evaluation}. Fitts' Law has also been applied effectively in various studies to analyze the performance of selecting targets (e.g. \cite{crossman1983feedback}, \cite{keele1968processing}). Many extensions of the original Fitts' Law have been proposed for different case scenarios. For instance, MacKenzie \emph{et al.} proposed an extension from a one-dimension to a 2D model for target acquisition tasks enabling the improvement of the index of difficulty for interactive computer systems with higher accuracy \cite{mackenzie1992extending}. 

Fitts' Law is a popular tool for user studies that has been extensively applied in evaluations \cite{gori2017one}. For example, Kim \textit{et al.} analyzed the usability of touch-key sizes in a driving safety simulation \cite{kim2014effect}. Hansen \textit{et al.} studied the performance of gaze and head tracking for point and selection tasks on head-mounted displays (HMDs) \cite{hansen2018fitts}. In addition, researchers applied Fitts' Law to reduce dwell-time for gaze-based interactions by taking into account the estimated target acquisition time and eye movement time \cite{isomoto2018dwell}. 

\subsection{Eye-strain Models}
Researchers have proposed various means to measure eye-strain based on eye movement analysis. Lanthier \textit{et al.} showed that eye fixations and eye-strain increases with fatigue \cite{lanthier2013measuring}. Komogortsev \emph{et al.} proposed the Fixation Quantitative Score (FQnS) to consider the number of fixation points in regards to a stimulus, and Fixation Qualitative Score (FQlS) based on distance drift of ﬁxation points that may reveal physical eye-strain \cite{komogortsev2010standardization}. Furthermore, Vasiljevas \textit{et al.} adopted an analytical model for muscle fatigue to assess eye-strain in gaze-based tasks \cite{vasiljevas2016modelling}. Researchers have also applied self-evaluation rating questionnaires to measure eye-strain for gaze-based applications \cite{majaranta2009fast}. Further, saccades-based approaches were proposed to measure eye-strain \cite{bahill1975overlapping, megaw1983visual, di2013saccadic}. However, analysis of saccades might raise challenges and need extra processing on budget-friendly devices, and therefore fixation-based approaches have been preferred \cite{abdulin2015user}. 

Considering the previous works, we propose a dual-purpose approach that can be applied in user studies with eye trackers to measure eye-strain based on screen and target properties for a specific duration. In our previous works, we proposed one approach to measure eye-strain involving subjective ratings (FELiX) \cite{felix} and introduced an index of difficulty for eye-tracking applications (IDEA) \cite{IDEA}. These approaches were compound models based on subjective and objective measures. The introduction of ESPiM is based on objective measures only, which may improve and optimize user interface design concepts based on eye-strain criteria. Based on the results of our previous models, we propose a predictable objective model suitable for assessments of visual prototypes on digital displays being used for a specific time to reduce costs of productions by comparing design ideas in early steps via pilot studies.

\section{Eye-Strain Probation Model (ESPiM)}

\begin{figure}[ht]
\centering
\includegraphics[width=0.23\textwidth]{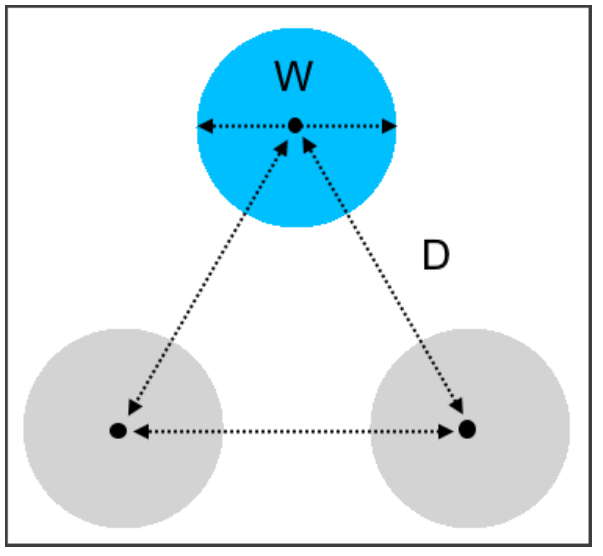}
\caption[Overview of the spatial parameters of ESPiM.]{Screenshot of the `FittsStudy' application \cite{wobbrock2011effects} with 3 targets. The blue target represents the active target to be selected. The distance between target centers is shown with D, and the width of targets is shown with W.}
\label{espim_overview}
\end{figure}

The ESPiM model relies on spatiotemporal parameters related to the screen size, target dimensions and distances (spatial) shown in Figure \ref{espim_overview}, task duration time (temporal), and eye-tracking fixations as described below:

\begin{enumerate}
    \item x: width of the screen in pixels.\\
    $x \in \mathbb{R}_{} \land x > 0$

    \item y: height of the screen in pixels.\\
    $y \in \mathbb{R}_{} \land y > 0$

    \item z: diagonal length of the screen in pixels. \\
    $z = \sqrt{x^{2} + y^{2}} \land z \in \mathbb{R}_{} \land z > 0$
    
    \item Area of screen (AoS): the surface area of the screen in which test applications are executed on in pixels.\\
    $AoS = x \times y \land AoS \in \mathbb{R}_{} \land AoS > 0$
    
    \item Area of target (AoT): the surface area of the target (user interface element) in which the user tries to focus on in pixels. Typically there are multiple targets on a user interface. We calculate the area of a single target since users focus on one target at selection time. In the case of targets with various areas, the average of all targets will be considered. \\
   $AoT \in \mathbb{R}_{} \land 0 < AoT \leq AoS$
    
    \item Distance of target (D): distance of the target centers from each other in pixels which must be smaller than or equal to screen's diagonal length.\\
    $D \in \mathbb{R}_{} \land 0 < D \leq z$
    
    \item Width of target (W): width of target on the screen in pixels, which must be smaller than or equal to screen width. \\
   $W \in \mathbb{R}_{} \land 0 < W \leq x$
    
    \item Shannon code ($\log_{2}(1 + \frac{D}{W})$): index of difficulty for point-and-select tasks \cite{mackenzie1989note} based on the Fitts' Law \cite{fitts1954information}. \\
    $\log_{2}(1 + \frac{D}{W}) \in \mathbb{R}_{} \land \log_{2}(1 + \frac{D}{W}) > 0$
    
    \item Task duration (TD): time spent on a specific task. \\
    $TD \in \mathbb{R}_{} \land TD > 0$
    
    \item Average number of fixations (ANF): average number of fixations recorded by an eye-tracking sensor of an entire task. \\
    $ANF \in \mathbb{R}_{} \land ANF > 0$
    
\end{enumerate}

\noindent
ESPiM: the calculated eye-strain value is based on Equation \ref{espim_equation} which is greater than 0 based on the square root function growth: ($ESPiM \in \mathbb{R}_{} \land ESPiM > 0 $).

The ESPiM model is based on pure test conditions such as screen and target properties regarding the dimensions and distances to be measured for a desired duration of time. This model provides an initial assessment to researchers about task difficulties. The ESPiM model can be used considering any 2D flat display type such as smartphones, tablets, and laptop/desktop monitors. This enables researchers to predict the difficulty level of target selection tasks on any device regardless of the applied interaction techniques. 

The ESPiM model given in Equation \ref{espim_equation} reflects \textit{the level of difficulty given the size and distances of targets over the screen} ($\log_{2}(1 + \frac{D}{W})$), \textit{to select a portion of the screen covered by a target} ($\frac{AoS}{AoT}$), \textit{considering an amount of eye fixation points to focus on targets} ($ANF$) \textit{for the specific period of time} (TD). The equation is offset by 1 in the case of minimal values for the parameters described earlier. The purpose of this addition is to set the minimum threshold of the square root function to start from $\sqrt{ \frac{\epsilon + 1}{ \lambda + 1}}$, where $\epsilon$ and $\lambda$ denote very small values. The average number of fixations (ANF) is one of the most used eye-tracking variables, which contributes to the accuracy of the calculations. 

\begin{figure}[ht]
	\centering
		\includegraphics[width=0.50\textwidth]{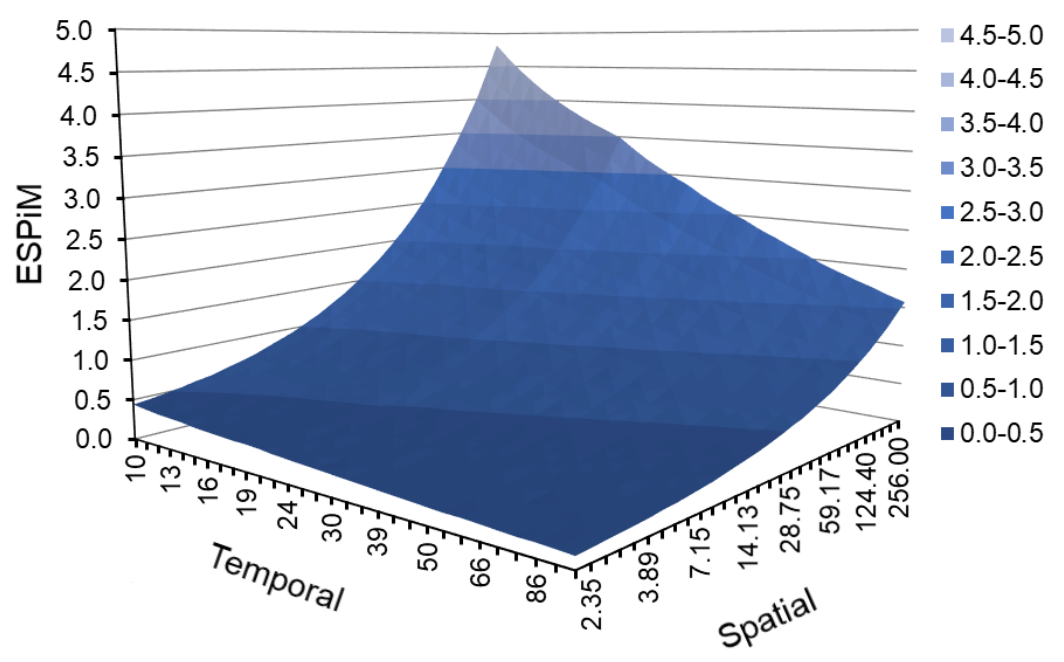}
	\caption[A 3D visualization of the ESPiM model.]{A 3D visualization of the ESPiM model for sample generated values. The spatial parameter axis represents the product of the relative area ratio of the screen over target, and the Shannon code as described in Equation \ref{espim_equation} with a constant average number of fixations (ANF).}
	\label{3d_visualization}
\end{figure}

These parameters are bound into the square root function to (1) shape the equation outcome into continuous values that are suitable for machine learning algorithms, and (2) downscale possible large numbers due to multiplications of parameters. Since the \textit{time} unit of denominator is integrated into the equation under the square root function, it may be ignored, and therefore we assign the unit of \textit{bits} for the ESPiM model for simplicity of comparisons. Figure \ref{3d_visualization} shows a 3D visualization of the ESPiM model calculated for sample generated values.

\begin{equation}
\label{espim_equation}
ESPiM =\sqrt{\frac{\big(~~\overbrace{~~(\frac{AoS}{AoT})\times\log_{2}(1+\frac{D}{W}}^{spatial})~~\times\overbrace{ANF}^{eye-tracking}\big) + ~~ 1}{\underbrace{TD}_{temporal}+1}}
\end{equation}

\subsection{Applications of ESPiM}
\label{applications-of-ESPiM}

Although we have primarily designed the ESPiM model to measure eye-strain on digital displays, it can be applied to (a) compare user interface designs, (b) compare different display devices, and (c) compare different interaction techniques based on eye-strain of users. As the spatial parameters are included in the ESPiM model, it can be used to estimate eye-strain before testing in a user study which can be beneficial to both research communities and industrial producers of digital content. 

In fact, we incorporate spatial parameters of screen and targets including applying Fitts' Law and eye-tracking fixation points for a desired period in a single measure. Since eye fixations are typically bound to a specific range (200-600~ms) \cite{majaranta2014eye}, the average number of fixations (ANF) for a desired time can be estimated. This property makes ESPiM a suitable choice for testing and evaluating interfaces even when there is no access to eye-tracking sensors to record eye fixations, since this parameter can be estimated based on the finite range of ANF for a desired test duration. ESPiM can be considered as an \textit{extension} to the Fitts' Law regarding eye-tracking case scenarios. 

\section{User Study 1: Fitts' Study (in-person)}

We conducted two user studies to evaluate the ESPiM model using (1) an in-person infrared desktop eye tracker study and (2) a video-based eye-tracking remote study. The purpose of the first evaluation was to study the ESPiM model. In this study, we used the unpublished datasets of our previous paper EyeTAP \cite{EyeTAP} in which we collected large amounts of infrared eye-tracking data. 

\subsection{Methods}
During this study, participants (19~Male, 13~Female) with an average age of 25.96 years ($SE=0.53$~years), were asked to select circular targets using eye gaze, and specifically the dwell-time method. We used the `FittsStudy' V4.2.7 application \cite{wobbrock2011effects} for our user study, which enabled us to run experiments based on Fitts' Law. Our stimuli contain two target widths (96 and 128 pixels) at three distances apart (256, 384 and 512 pixels) to record the required measures. An activation of 500~milliseconds for dwell-time was used as this is the best-suited threshold in previous studies \cite{mackenzie2012evaluating, vspakov2004line}. Thus, when a participant focuses for 500~milliseconds on a target, it triggers a click event, and the target gets selected, and any gaze movement from the target borders causes pointer movement and therefore restarts the target selection process. 

The user study took 12 minutes on average for each participant, 10 minutes for preparation, including description of the task, training and eye calibration, and 2 minutes for the actual target selection task in 6 trials (2 widths $\times$ 3 distances). Target selection using an eye-tracking sensor with a low activation threshold (500~ms) challenges the control of the pointer on the screen. Conventional target selection input devices such as keyboard and mouse enable users to select targets at their pace. Considering the differences between these selection methods, the relatively short duration of target selection in our study using eye-tracking was sufficient to record required measures as well as not to overwhelm participants with heavy tasks. 

To capture the eye-tracking data, a remote eye-tracking sensor (Tobii 4C) with a sampling rate of 90~Hz on a monitor with the resolution of $1920 \times 1080$ pixels (24$''$) running on an Intel i7 Windows 10 PC was used. Subjects sat with a distance of 60~cm ($\approx$23.5 in) to the eye tracker. The following data was collected during the study: (1) movement time based on the Fitts' Law, (2) recorded errors of target selections, (3) average number of fixations (ANF), and (4) Fixation Qualitative Score (FQlS) (a measure that can reveal physical eye fatigue based on distance drift of ﬁxation points) \cite{komogortsev2010standardization}. These measures accompanied by ESPiM calculations based on Equation \ref{espim_equation} were needed for our analysis.

We informed participants about the procedures of the studies, and registered their consents before conducting the experiments.

\subsection{Results and Discussion}

\begin{figure*}[ht]
   \subfigure{\includegraphics[width=0.17\textwidth]{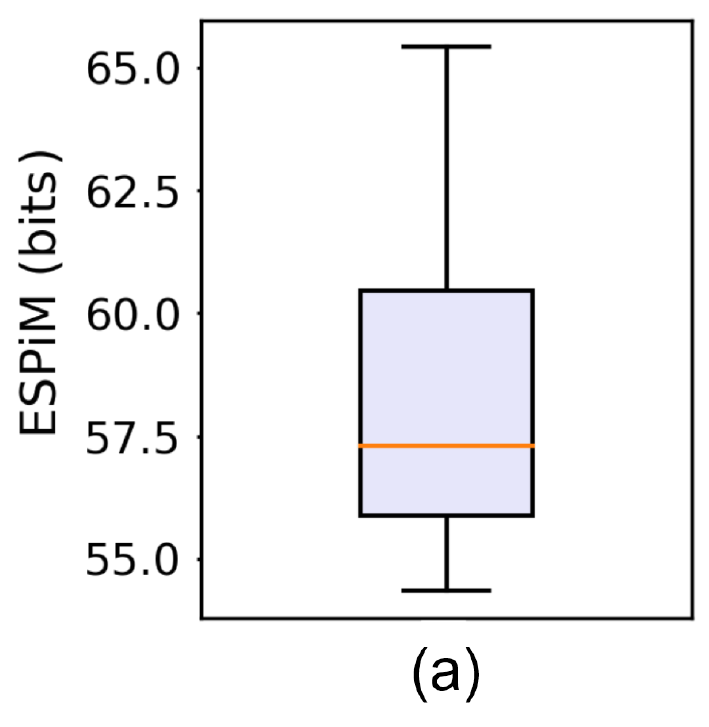}
		\label{total_espim}}
    \subfigure{\includegraphics[width=0.17\textwidth]{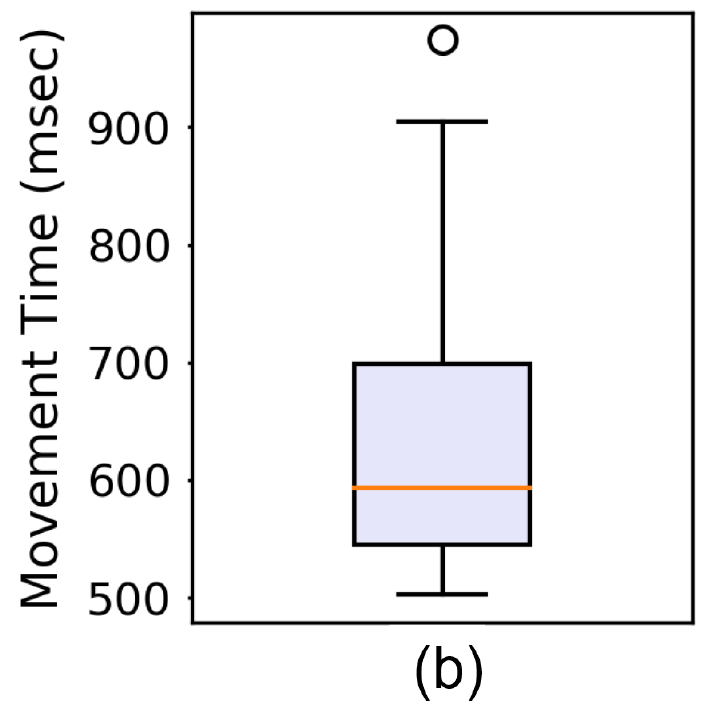}
		\label{total_mt}}
    \subfigure{\includegraphics[width=0.17\textwidth]{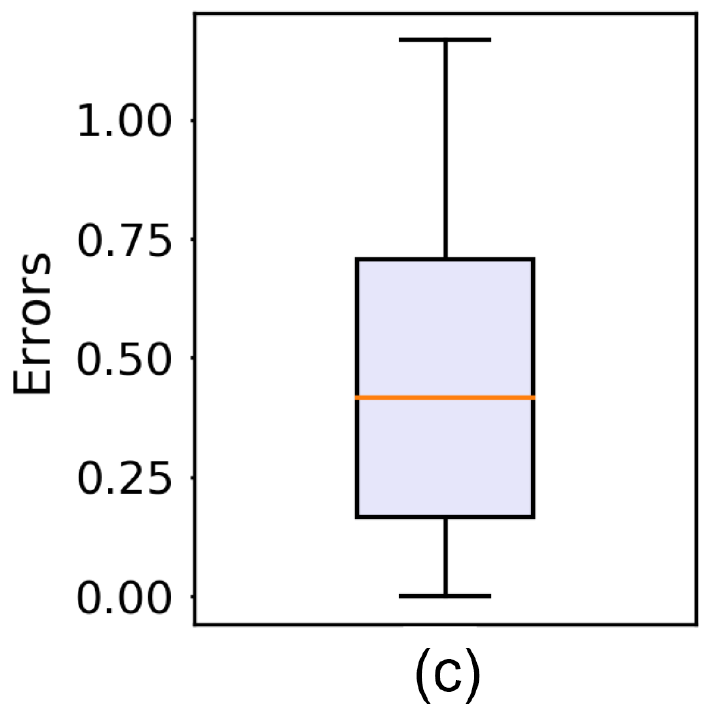}
		\label{total_er}}
  \subfigure{\includegraphics[width=0.17\textwidth]{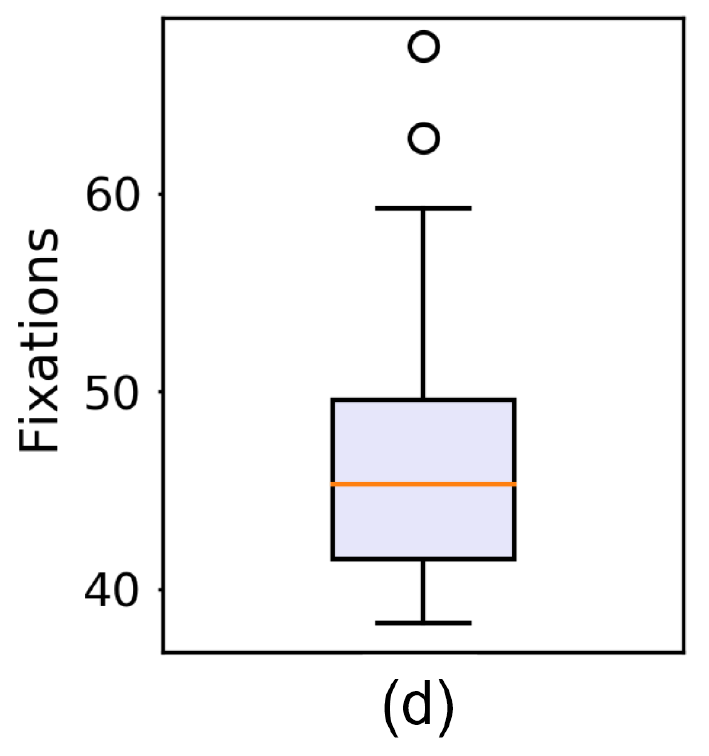}
		\label{total_fixations}}
    \subfigure{\includegraphics[width=0.17\textwidth]{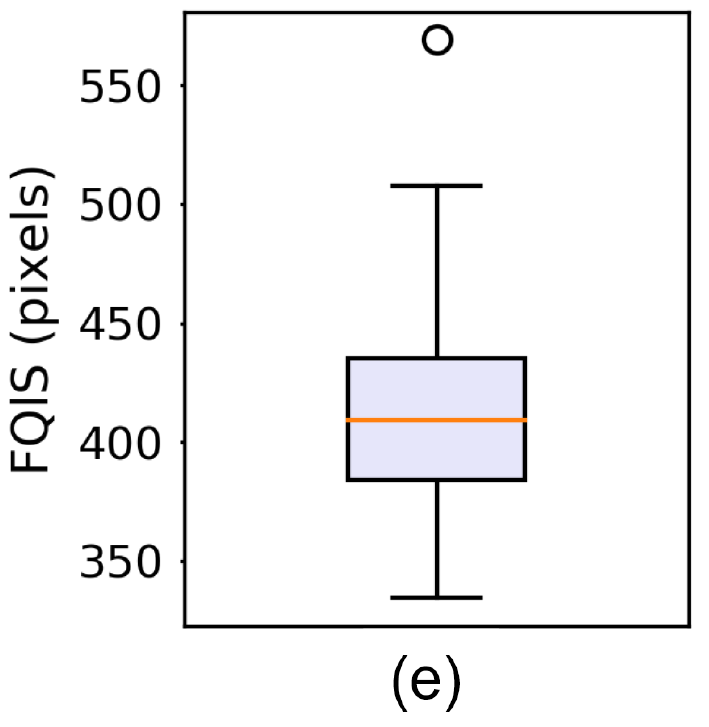}
		\label{total_fqls}}
  \caption[ESPiM results of the first evaluations.]{Illustration of (a) ESPiM, (b) calculated movement time based on Fitts' Law, (c) recorded errors, (d) eye fixations, and (e) Fixation Qualitative Score (FQlS) measure for 32 participants observed in our study.}
\label{total_charts}
\end{figure*}

The results of our experiments were analyzed using the JASP software \cite{JASP2019}. We analyzed our raw data based on three categories: (1) model analysis regarding the effectiveness of ESPiM, and (2) participants' frequency of video gameplay. Figure \ref{total_charts} illustrates the calculated ESPiM values and the recorded measures. ESPiM can be applied to compare the results of different groups in user studies. We analyze these basic results among groups of video gameplay frequency.  

\begin{figure}[ht]
\centering
\includegraphics[width=0.33\textwidth]{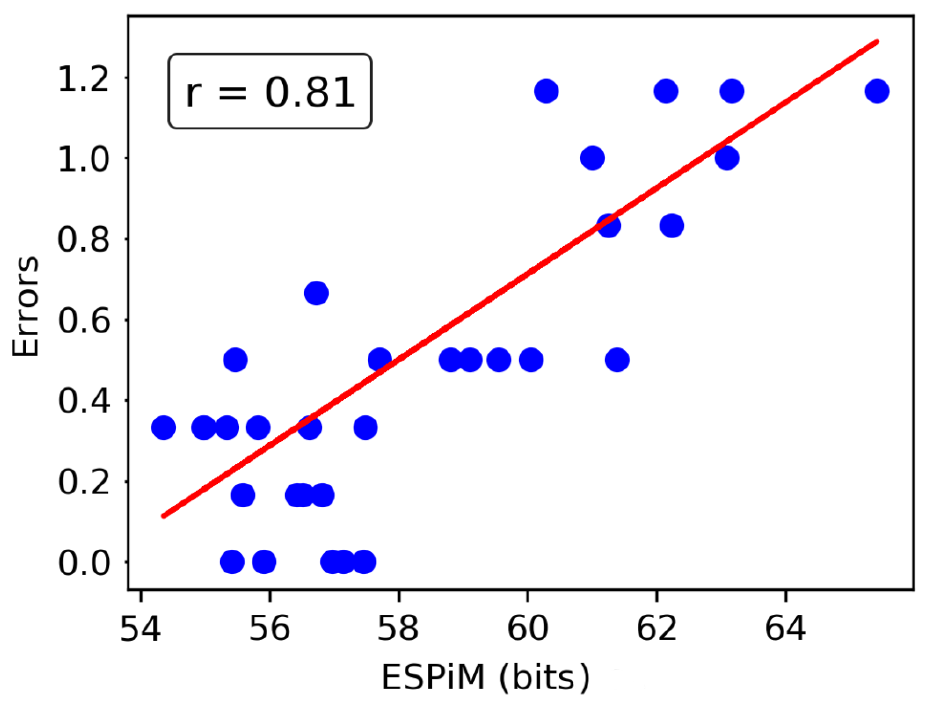}
\caption[Overview of ESPiM and error rates.]{Correlation of errors and ESPiM (Pearson's r = 0.81, \textit{p} $<$ .001).}
\label{espim_errors_correlation}
\end{figure}

We found a strong correlation (Pearson's r = 0.81, and \textit{p} $<$ .001) between the measured ESPiM values and the recorded errors during the user study. This suggests a higher error rate for higher eye-strain values as illustrated in Figure \ref{espim_errors_correlation}.

\subsubsection{Video-gameplay Analysis}

\begin{figure}[ht]
\centering
\includegraphics[width=0.33\textwidth]{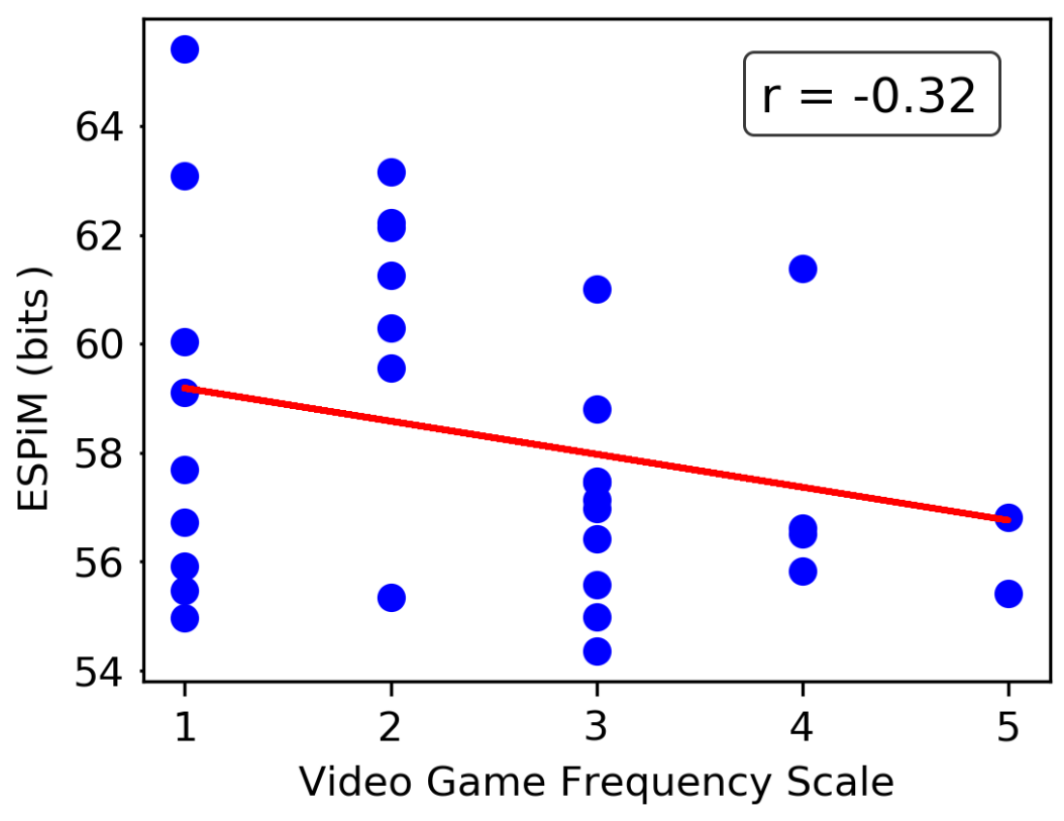}
\caption[Test results of the video gameplay groups.]{Correlation of video game frequency and the calculated ESPiM values (Pearson's $r=-.32$).}
\label{video_espim_correlation}
\end{figure}

In a pre-test questionnaire, we asked the participants to rate their video gameplay frequency on an integer scale of 1 (never) to 5 (every day). We found a negative correlation (Pearson's r = -0.32, and \textit{p} $>$ .05) which was not statistically significant, although it shows a trend for further discussion as shown in Figure \ref{video_espim_correlation}. In order to further analyze the result, we divided our subjects into two groups based on their rate regarding the mean (2.5) of the scale. Participants with a rate lower than the average value are labeled as \textit{low} and those with a rate greater or equal than the average as \textit{high} groups. 

Each group was assigned exactly 16 participants. In general, we found significantly higher eye-strain level, movement time, and error rates for the low group compared to the high group, while no significantly different values for the eye fixations and the FQlS measures were recorded. 

\begin{figure*}[ht]
  \centering
  \subfigure{\includegraphics[width=0.18\textwidth]{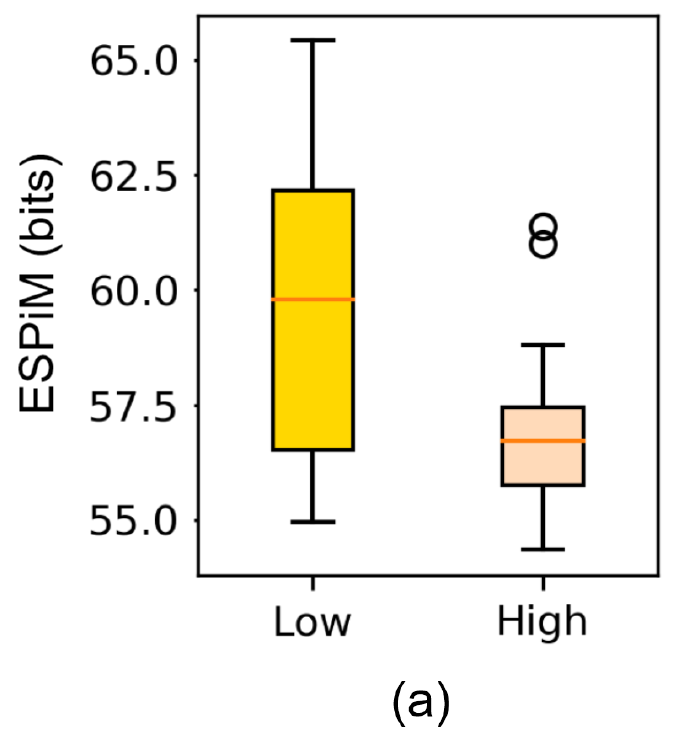}
		\label{video_espim}}
    \subfigure{\includegraphics[width=0.18\textwidth]{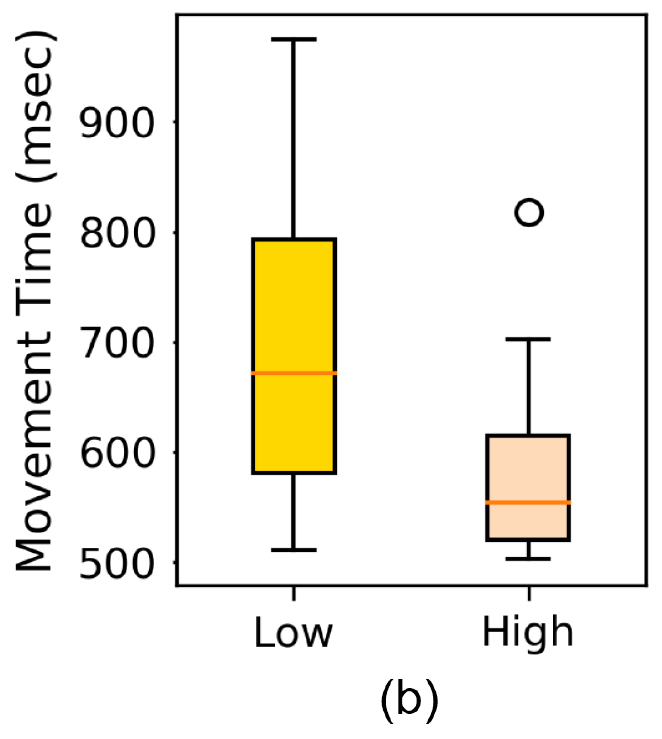}
		\label{video_mt}}
    \subfigure{\includegraphics[width=0.18\textwidth]{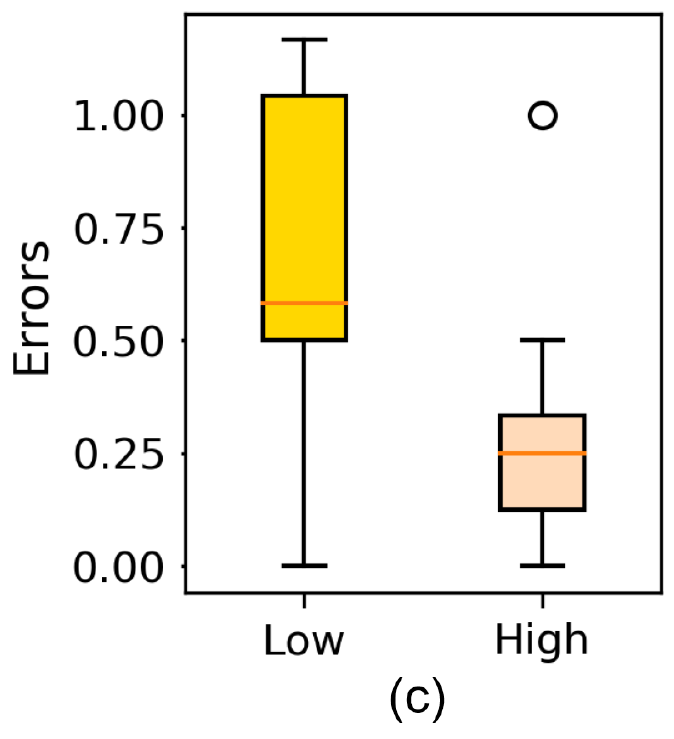}
		\label{video_er}}
  \subfigure{\includegraphics[width=0.17\textwidth]{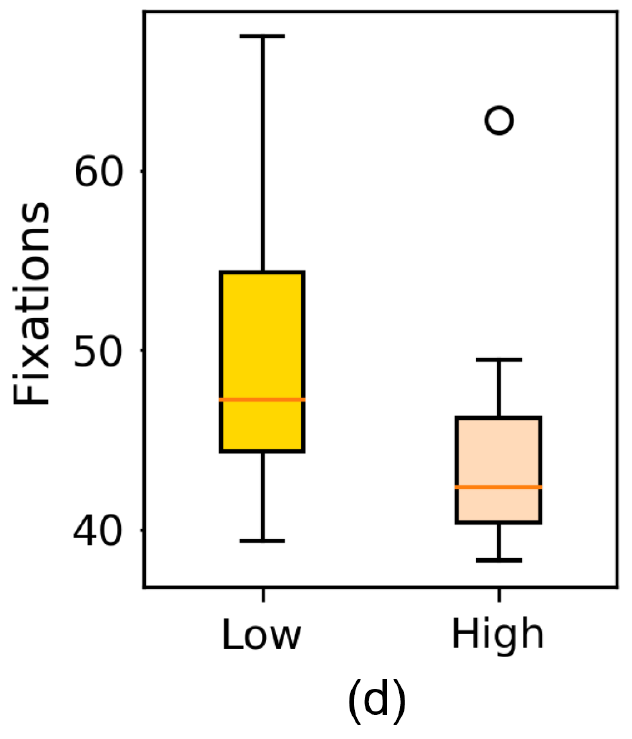}
		\label{video_fixations}}
  \subfigure{\includegraphics[width=0.18\textwidth]{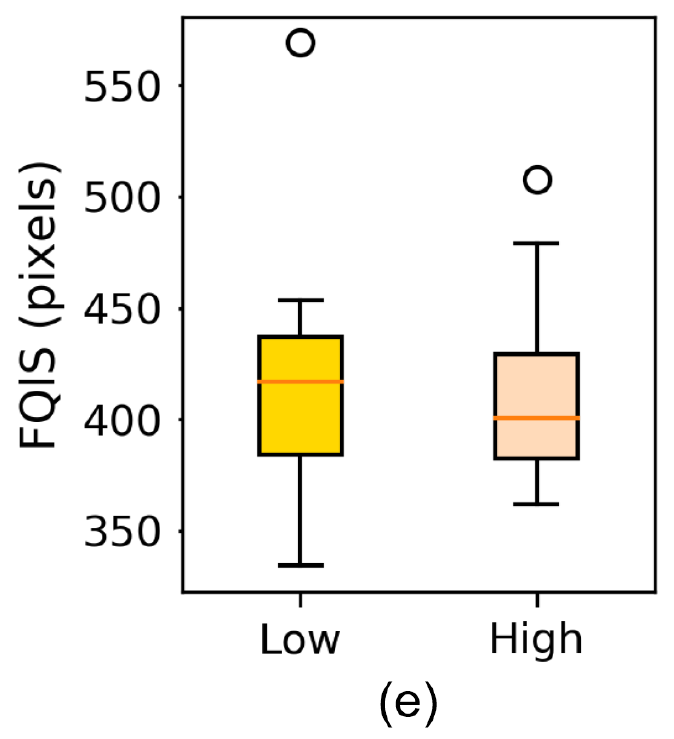}
		\label{video_fqls}}
  \caption[Illustration of various evaluation parameters.]{Illustration of (a) ESPiM ($\textit{p} < .05$), (b) movement time based on Fitts' Law ($\textit{p} < .05$), (c) error rates ($\textit{p} < .05$), (d) eye fixations, and (e) Fixation Qualitative Score (FQlS) ($\textit{p} > .05$) based on video gameplay frequency of 32 participants divided into 2 equal groups (Low, High) with 16 participants.}
\label{video_frequncy_charts}
\end{figure*}

\noindent \\
\textbf{Eye-strain Probation Model (ESPiM):} We applied a paired-sample t-test to check the effect of video gameplay frequency on the calculated ESPiM  ($t(15)=2.14, \textit{p} < .05$) and found a significant difference between low ($M=59.52~bits$, $SE=0.81~bits$) and high ($M=57.04~bits$, $SE=0.48~bits$) groups as illustrated in Figure \ref{video_espim}. Thus, those that frequently play video games experienced a significantly lower amount of eye-strain than the group of lower frequency of playing video games. \\

\noindent
\textbf{Movement Time:} We applied a paired-sample t-test to check the effect of video gameplay frequency on movement time based on the Fitts' Law and found a significant difference ($t(15)=2.36, \textit{p} < .05$) between low ($M=690.37~msec$, $SE=34.63~msec$) and high ($M=584.72~msec$, $SE=21.53~msec$) groups as illustrated in Figure \ref{video_mt}. Thus, those with a higher frequency of video gameplay were able to perform the task in a shorter amount of time than those who play video games less frequently. \\

\noindent
\textbf{Error Rates:} We applied a paired-sample t-test to check the effect of video gameplay frequency on the recorded errors and found a significant difference ($t(15)=3.62, \textit{p} < .05$) between low ($M=0.69~errors$, $SE=0.09~errors$) and high ($M=0.27~errors$, $SE=0.06~errors$) groups as illustrated in Figure \ref{video_er}. This suggests that those with a higher frequency of video gameplay made less errors than those who do not play or play less frequently. \\

\noindent
\textbf{Eye Fixations:} We applied a paired-sample t-test to check the effect of video gameplay frequency on eye fixations with ($t(15)=2.12, \textit{p} > .05$) and found no significant difference between low ($M=49.53~fixations$, $SE=1.93~fixations$) and high ($M=44.13~fixations$, $SE=1.51~fixations$) groups as illustrated in Figure \ref{video_fixations}. Thus, gameplay frequency did not impact eye fixations during the test.  \\

\noindent
\textbf{Fixation Qualitative Score (FQlS):} We applied a paired-sample t-test to check the effect of video gameplay frequency on the calculated FQlS measure (a measure that can reveal physical eye fatigue based on distance drift of ﬁxation points) with ($t(15)=0.28, \textit{p} > .05$) and found no significant difference between low ($M=416.00~pixels$, $SE=13.13~pixels$) and high ($M=411.67~pixels$, $SE=10.12~pixels$) groups as illustrated in Figure \ref{video_fqls}. In other words, gameplay frequency did not impact fixation drift distances.

We found a similar pattern on higher ANF (Average Number of Fixations) measure ($IQR_{low} > IQR_{high}$, and ~$Range_{low} > Range_{high}$). However, there were no significant differences among low and high groups, the relatively higher ANF was the cause of the higher ESPiM values for the \textit{low} group. This suggests that participants with higher gameplay frequency were more experienced in moving their eyes in a shorter time and therefore produced lower fixations and consequently lower eye-strain based on our model. 

This result may be due to the fact that video games can increase visual abilities. Previous studies have shown the relationship between video gameplay and eye movements, for instance, shorter saccadic reaction time in video game players \cite{MACK201426}, the usage of video games to train visual skills \cite{achtman2008video}, and to enhance visual search in players \cite{CASTEL2005217}. These characteristics of frequent game players enable them to experience lower eye-strain, as indicated by our results.

We observed that participants with a high video gameplay frequency produced lower errors than their counterparts in target selections as shown in Figure \ref{video_er}. The cause of low error rates in participants with high video game experience is due to their fast eye movements, as shown in Figure \ref{video_mt}. Thus those who have more ``trained'' visual search through video gameplay\cite{achtman2008video} tend to perform better with eye-tracking-based selection. 

The mentioned observations based on the higher performance of experienced participants related to video gameplay cannot be interpreted as a promotion in favour of video games.

\section{User Study 2: Focus Shift Simulator (online)}

The second study, described here, was conducted to include a more in-depth analysis of eye-strain characteristics using a video-based eye-tracking technique applied via the WebGazer API \cite{papoutsaki2016webgazer}, a recognized tool for remote eye-tracking user studies.

\subsection{Methods}
We developed a custom web application that runs on the client-side completely and requires no video footage to be sent to the server. The WebGazer API \cite{papoutsaki2016webgazer} uses the user's webcam to track eye movement and maps features of the eye and positions on the screen. It begins by obtaining the participant's consent to use the webcam, followed by a calibration process. During this process, the user has to click on reference points with the mouse while looking at the cursor. 

Although the eye-tracking accuracy of the WebGazer API relies on the amount of light in the environment, we informed participants to run the application in a well-lit environment. In addition, the WebGazer API applies a learning mechanism from user interactions and contains a self-calibration functionality to improve its accuracy and therefore provides a sufficient approximation of the user's gaze points \cite{papoutsaki2016webgazer}.

\begin{figure}[ht]
	\centering
		\includegraphics[width=0.45\textwidth]{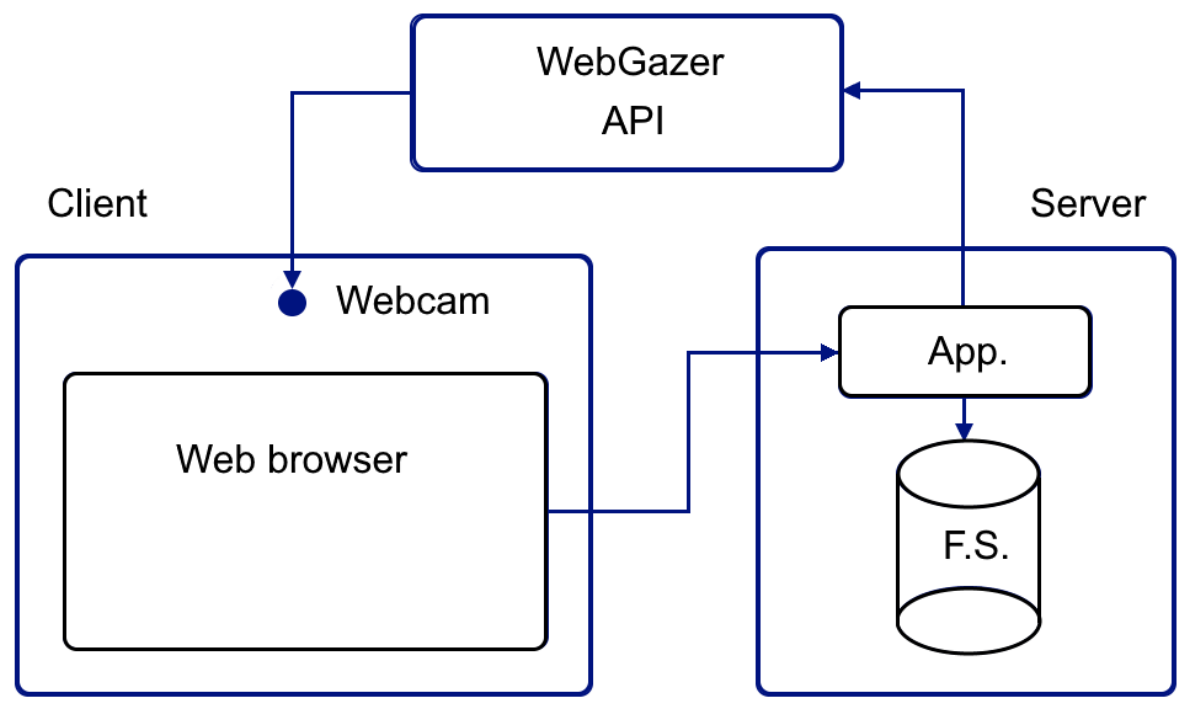}
	\caption[Overview of the focus shift simulator application.]{Overview of the focus shift simulator test application using the WebGazer API \cite{papoutsaki2016webgazer}. The user runs the application from a web browser, the client-server connection is established, and the WebGazer API acquires control of the webcam to track the user's eyes. The results of the entire test session is saved on the file system (F.S.) on the server.}
	\label{focus_shift_simulator_pipeline}
\end{figure}

The application was developed in \textit{Java} and the server interaction with the WebGazer API \cite{papoutsaki2016webgazer} was implemented in \textit{JavaScript}. The AWS Elastic Beanstalk \cite{aws_Elastic} provided by Amazon Web Services was used for deploying our web application. Figure \ref{focus_shift_simulator_pipeline} shows the workflow of the \textit{focus shift simulator} user study.  

In terms of the study, participants were asked to launch the developed web application at 3 times in the day (from 08:00 to 12:00, 12:00 to 18:00 and after 18:00) over 7 days (did not have to be consecutive). Prior to each trial in the study we asked the users how many hours they have been working on or using a digital display. 

\begin{figure}[ht]
\centering
\includegraphics[width=0.55\textwidth]{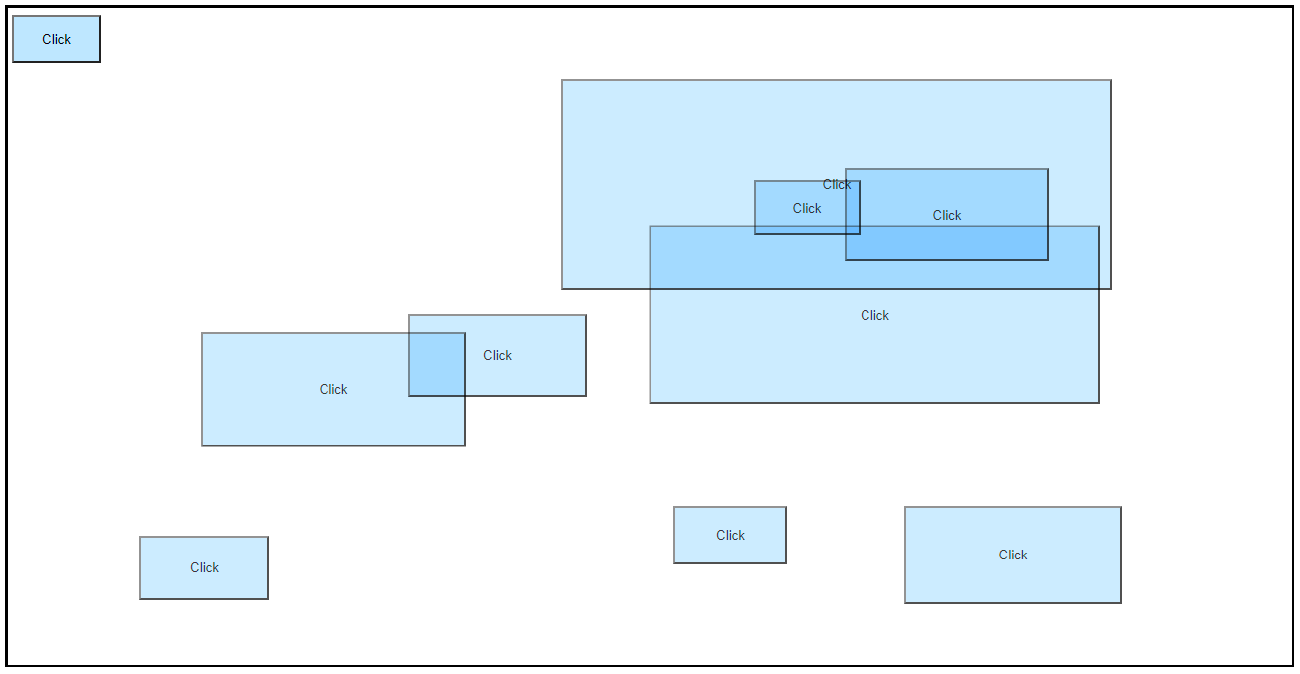}
\caption[Focus shift simulator.]{The screenshot of the focus shift simulator application for 10 targets. We reduced the number of targets for this screenshot to reduce overlapping of targets for higher visibility. Targets contain `click' label and appear one by one randomly across the screen with different sizes. As soon as user clicks on a button, the application removes the selected button and loads the next.}
\label{screenshot}
\end{figure}

Next, the participant was prompted to click on 30 buttons that randomly appeared on the screen one at a time. Each button has a different position and size adjusted to the screen dimension as illustrated in Figure \ref{screenshot}. By clicking on the last button, the recorded raw data is sent to the server. At the end of the trial, we asked the participant to rate their current eye-strain level on a scale from 1 (none) to 5 (a lot).

The intuition behind the developed \textit{focus shift} application is based on the multi-tasking efforts to interact with different interactions, applications, and events while working on a digital display. The users need to move their focal points to react to situations. For instance, reading emails and constantly receiving messages on a messenger application and monitoring running applications in the background, which may interrupt the user's attention with prompt dialog boxes to confirm/decline certain actions are typical tasks of an office employee. 

\begin{figure}[ht]
\centering
\includegraphics[width=0.55\textwidth]{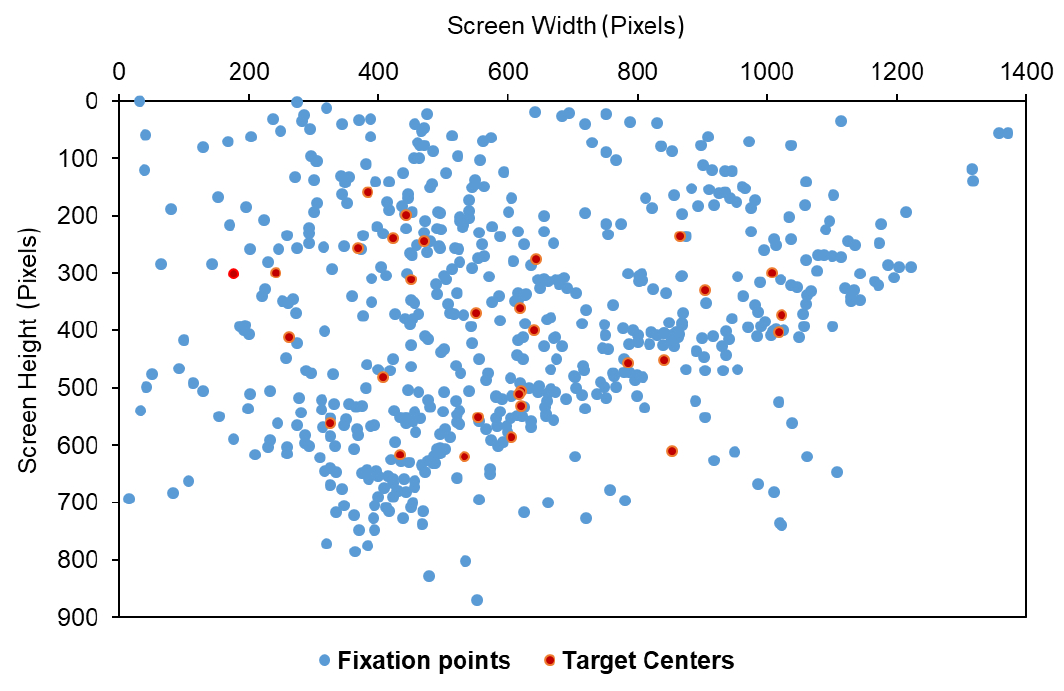}
 \caption[Fixation points on screen.]{The recorded fixation points of an entire test session with 30 targets on a laptop display. The red circles represent center of targets, and the blue circles represent fixation points. This figure shows the intuition of fixation points to be integrated in our eye-strain model to be considered for a specific period of time.}
\label{fixation_scatter}
\end{figure}

Figure \ref{screenshot} illustrates the screenshot of the test application to simulate multi-tasking events on a computer and Figure \ref{fixation_scatter} shows the pattern of eye fixations when shifting to different locations on the screen to follow stimuli and illustrates the intuition of fixation points integration in our proposed model (see Equation \ref{espim_equation}).

This study was specifically designed to be applied remotely to comply with physical restrictions caused by the COVID-19 pandemic \cite{covid_19_intro}. Moreover, since the study collects data from 24 hours a day from users, it could be executed anytime and anywhere via the URL to the server to provide participants freedom and control over their interactions with the system rather than running the tests in a laboratory equipped with eye-tracking sensors. Therefore the application of video-based eye-tracking was found to be the ideal choice of design.

After analyzing the preliminary results, we found that some participants failed to stick to the restricted time slots. This may be due to the fact that our testing sessions collided with personal tasks and that since the start of the COVID-19 pandemic \cite{covid_19_intro}, many workers began working from home with varying work routines rather than the traditional 9-to-5 workday. As we originally wanted to study our ESPiM model for classical 9-to-5 working hours, based on our data, we defined two groups of participants:

\begin{itemize}
    \item \textbf{9-to-5}: any time between 09:00 and 17:00 o'clock.
    \item \textbf{Flexible}: any times not in the 9-to-5 group.
\end{itemize}

These slots were chosen based on the concept of 9-to-5 working hours, and there are no intersections between the groups (9-to-5 $\cap$ Flexible = $\emptyset$). We had four hypotheses that we considered in this study:  

\begin{itemize}
    \item H1: Users perceive higher eye-strain level beyond the standard 9-to-5 working hours: ESPiM(9-to-5)~$<$~ESPiM(Flexible).

    \item H2: The group of 9-to-5 participants may cause more errors than the flexible group. In simple words, we assume that the flexible group may choose their preferred hours of work or take more time in between work periods and thus make fewer errors. 
    
    \item H3: The longer users spend on a digital display, the more eye-strain they have.
    
    \item H4: The increase of the calculated ESPiM value correlates with increased mouse pointer movements.
\end{itemize}

We collected 70 samples from 13 participants (10~Male, 3~Female) who partook in both experiments with an average age of 31.33 years ($SE=2.01$~years) to be analyzed based on their working times and ratings. Specifically, we collected (a) screen resolution, (b) test duration, (c) errors, (d) eye fixations, (e) display hours, (f) perceived eye-strain rating, and (g) eye-strain symptoms of participants.

We informed participants about the procedures of the studies, and registered their consents before conducting the experiments.

\subsection{Results and Discussion}

The results of our experiments were analyzed using the JASP software \cite{JASP2019}. Since the start of the COVID-19 pandemic \cite{covid_19_intro} many knowledge workers and students were forced to work from home, which typically deviates from the routine of 9-to-5 working hour schedules. This phenomenon motivated us to apply ESPiM to study the impacts of remote working via measuring eye-strain based on our proposed equation (see Equation \ref{espim_equation}). 

Although the calculated ESPiM values of both groups (9-to-5 and flexible) showed no significant difference, we observed significant patterns in the (a) subjective ratings, (b) time spent on a digital display, and (c) recorded error rates among groups.

\begin{figure*}[ht]
  \subfigure{\includegraphics[width=0.15\textwidth]{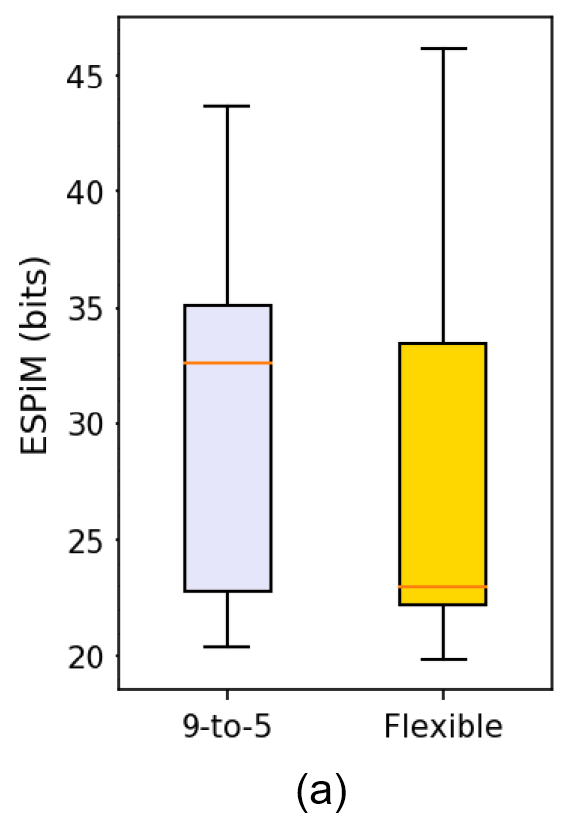}
		\label{espim_both_groups}}
  \subfigure{\includegraphics[width=0.15\textwidth]{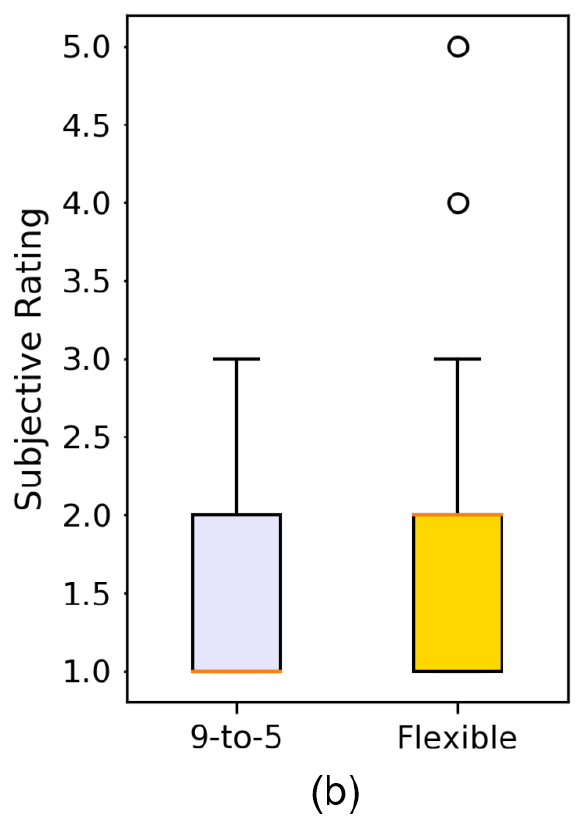}
		\label{ratings_both}}
  \subfigure{\includegraphics[width=0.15\textwidth]{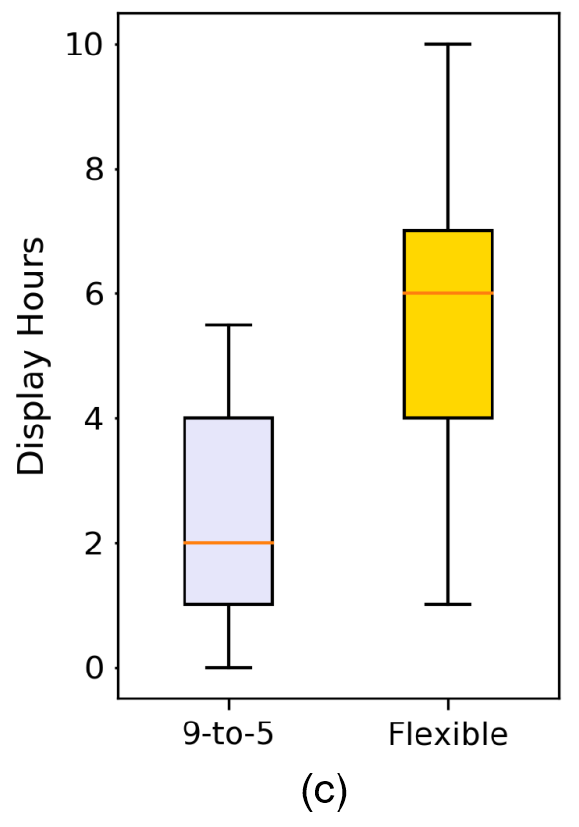}
    	\label{display_hours_both}}
  \subfigure{\includegraphics[width=0.16\textwidth]{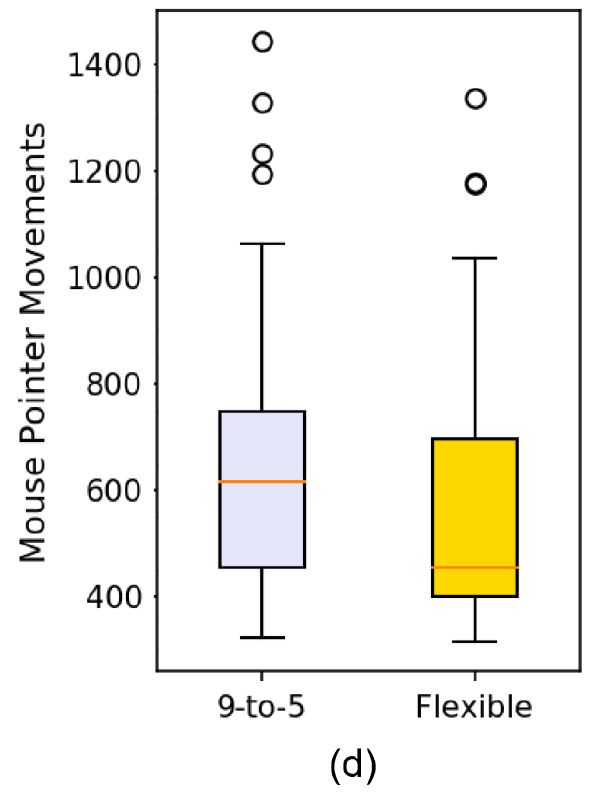}
		\label{mousemovement_both}}
  \subfigure{\includegraphics[width=0.15\textwidth]{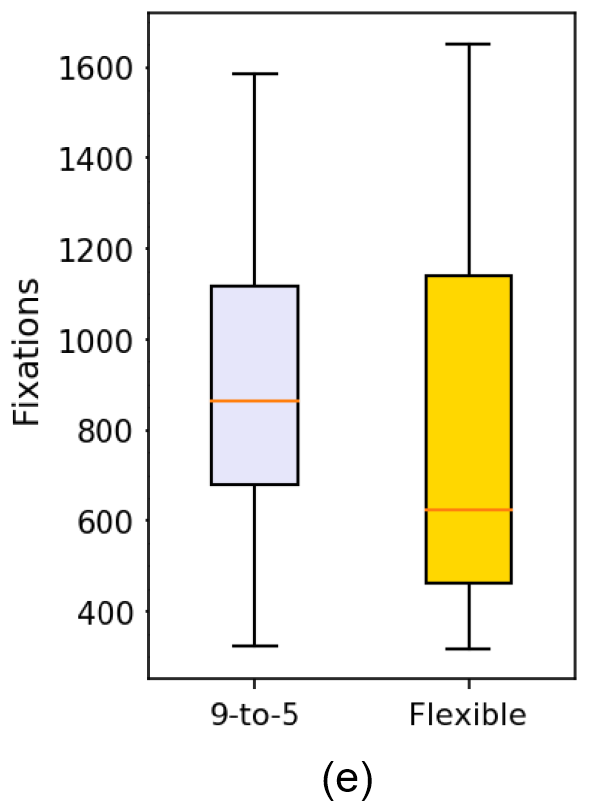}
		\label{fixations_both}}
  \subfigure{\includegraphics[width=0.15\textwidth]{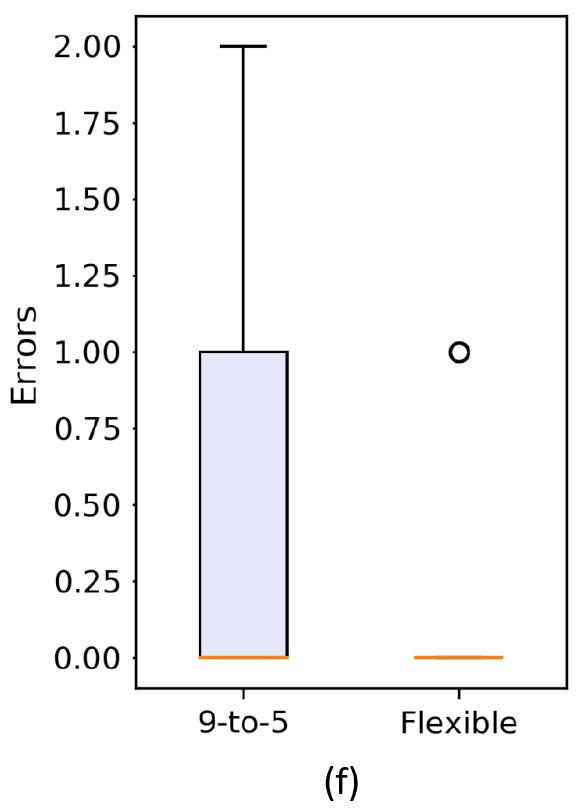}
		\label{errors_both}}
  \caption[ESPiM, rating and display hours results, mouse movements, fixations and error rates.]{Illustration of (a) ESPiM ({p} $>$ .05), (b) subjective ratings of perceived eye-strain level (\textit{p} $<$ .001), and (c) display hours before test (\textit{p} $<$ .001) of both testing groups, (d) mouse pointer movements of both testing groups (\textit{p} $>$ .05), (e) recorded eye fixation points (\textit{p} $>$ .05), and (f) error rates (\textit{p} $<$ .05).}
\label{espim_complete}
\end{figure*}

To test our first hypothesis, we applied a paired-samples t-test to check the difference of the calculated ESPiM (t(34)=1.90, \textit{p} $>$ .05) among both groups and found no significant difference between 9-to-5 ($M=30.58~bits$, $SE=1.22~bits$) and flexible ($M=27.76~bits$, $SE=1.34~bits$) groups as illustrated in Figure \ref{espim_both_groups}. This suggests our first hypothesis (H1) concerning a lower eye-strain level for 9-to-5 participants than the flexible group is rejected. 

To test our second hypothesis, we applied a paired-samples t-test to check the effect of errors on the calculated ESPiM (t(34)=2.47, \textit{p} $<$ .05) and found a significant difference between 9-to-5 ($M=0.31~errors$, $SE=0.09~errors$) and flexible ($M=0.08~errors$, $SE=0.04~errors$) groups as illustrated in Figure \ref{errors_both}. As we had predicted, there were higher error rates of the 9-to-5 group compared to the flexible group in our second hypothesis (H2), which is thus confirmed by these results. This shows that the flexible group, who are more flexible to work on their digital devices, could finish the test with fewer mistakes.

We applied a paired-samples t-test to check the effect of subjective ratings on the calculated ESPiM (t(34)=3.58, \textit{p} $<$ .001) and found a significant difference between 9-to-5 ($M=1.34~points$, $SE=0.10~points$) and flexible ($M=2.00~points$, $SE=0.15~points$) groups as illustrated in Figure \ref{ratings_both}. 

Further, we looked at the effect of display hours on the calculated ESPiM (t(34)=6.52, \textit{p} $<$ .001) and found a significant difference between 9-to-5 ($M=2.41~hours$, $SE=0.27~hours$) and flexible ($M=5.64~hours$, $SE=0.37~hours$) groups as illustrated in Figure \ref{display_hours_both}. 

Thus, although we predicted a higher eye-strain level based on ESPiM values for participants who spend more time on a digital display in our third hypothesis (H3), and we recorded significantly higher working hours for the flexible group than the 9-to-5 participants, the difference of ESPiM values was not statically significant and therefore we reject our third hypothesis. We posit that the reason for higher working hours but relatively similar eye-strain level of the flexible group lies in the fact that flexible participants could work anytime based on their convenience and therefore were benefited from breaks rather that those bound to a certain time window as the 9-to-5 group.

\begin{figure}[ht]
\centering
\includegraphics[width=0.40\textwidth]{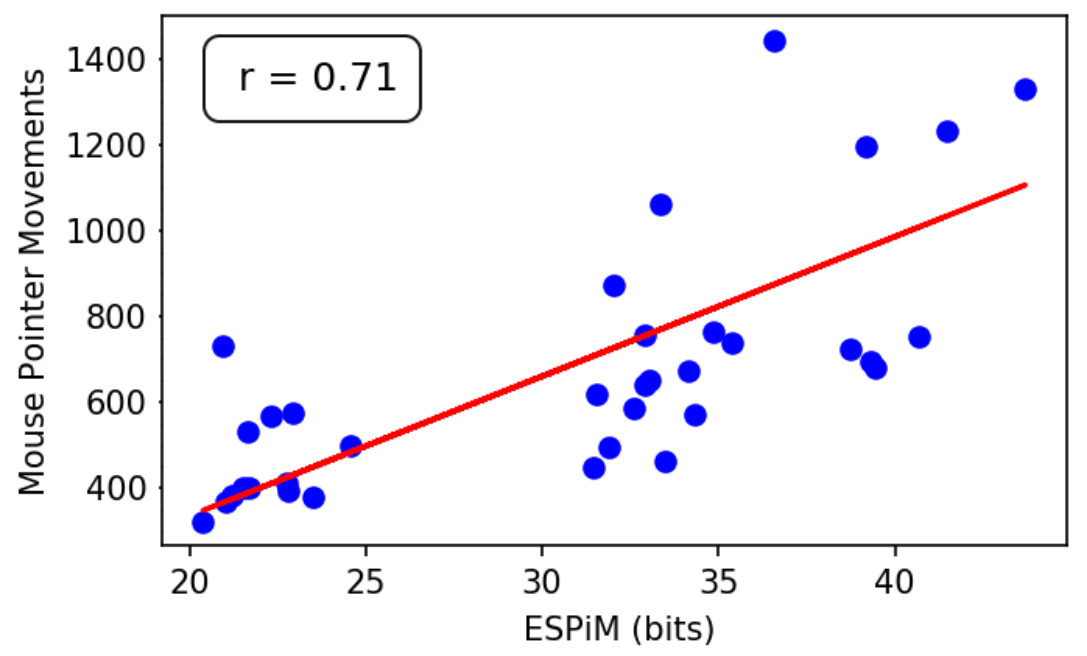}
\caption[Correlation of ESPiM and Mouse Pointer Movements 9-to-5.]{Correlations of ESPiM and mouse pointer movements for the \textbf{9-to-5} group (Pearson's $r=.71$, \textit{p} $<$ .001).}
\label{correlation_espim_mouse_9_to_5}
\end{figure}

\begin{figure}[ht]
\centering
\includegraphics[width=0.40\textwidth]{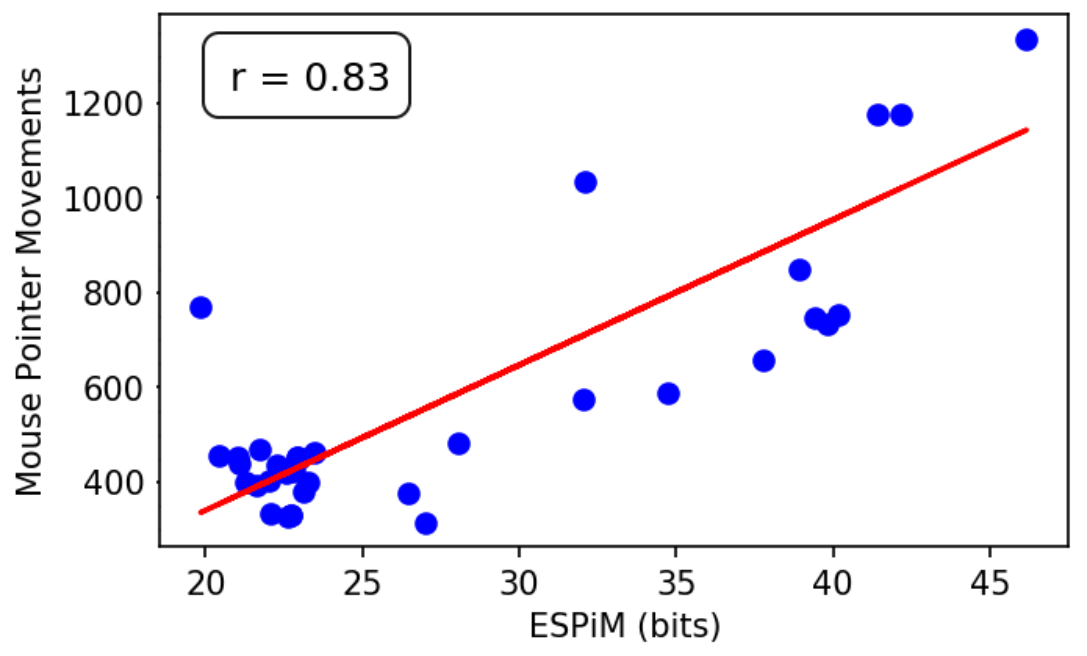}
\caption[Correlation of ESPiM and Mouse Pointer Movements flexible.]{Correlations of ESPiM and mouse pointer movements for the \textbf{flexible} group (Pearson's $r=.83$, \textit{p} $<$ .001).}
\label{correlation_espim_mouse_flexible}
\end{figure}

To test our fourth hypothesis, we studied the correlation of ESPiM and mouse pointer movements for each group and found positive correlations among those measures as predicted and therefore confirm the hypothesis (H4) as shown in Figure \ref{correlation_espim_mouse_9_to_5}, and \ref{correlation_espim_mouse_flexible}. This result suggests that tired eyes may lead to an increase of mouse pointer movements among users for target selection tasks. 

In addition, we studied the effect of mouse pointer movements (t(34)=1.81, \textit{p} $>$ .05) and found no significant difference between 9-to-5 ($M=667.14~movements$, $SE=47.87~movements$) and flexible ($M=565.11~movements$, $SE=45.04~movements$) groups as illustrated in Figure \ref{mousemovement_both}. However, the pointer movements correlate with the calculated ESPiM of each group. This shows that both groups of users applied a similar amount of mouse pointer movements during our test. Table \ref{tab:des_statistics_mouse_movements_rating} shows the details of the mouse pointer movement analysis.

\begin{table}[h!]
	\centering
	\caption[Descriptive statistics of mouse pointer movements.]{Descriptive statistics of mouse pointer movements.}
	\label{tab:des_statistics_mouse_movements_rating}
	{
		\begin{tabular}{|lcccc|}
						\hline
			& \multicolumn{2}{c}{Mouse Pointer Movements} & \multicolumn{2}{c|}{Subjective Rating} \\
			& 9-to-5 & Flexible & 9-to-5 & Flexible\\
			\hline
			Mean    &  667.14 & 565.11 & 1.34 & 2.00     \\
			Median  &  617.00 & 454.00 & 1.00 & 2.00    \\
			SD      &  283.22 & 266.50 & 0.59 & 0.93     \\
			IQR     &  291.00 & 295.50 & 1.00 & 1.00     \\
			Range   &  1121.00 & 1021.00 & 2.00 & 4.00   \\
			Min     &  322.00 & 315.00 & 1.00 & 1.00     \\
			Max     &  1443.00 & 1336.00 & 3.00 & 5.00   \\
			\hline
		\end{tabular}
	}
\end{table}

We also explored the effect of eye fixations on the calculated ESPiM (t(34)=1.46, \textit{p} $>$ .05) and found no significant difference between 9-to-5 ($M=904.28~fixations$, $SE=64.06~fixations$) and flexible ($M=791.08~fixations$, $SE=76.35~fixations$) groups as illustrated in Figure \ref{fixations_both}. This shows both groups of users experienced similar amount of eye fixations during our test. Further, the Figure \ref{errors_both} shows the recorded errors for both 9-to-5 and flexible groups.

\begin{figure}[ht]
\centering
\includegraphics[width=0.80\textwidth]{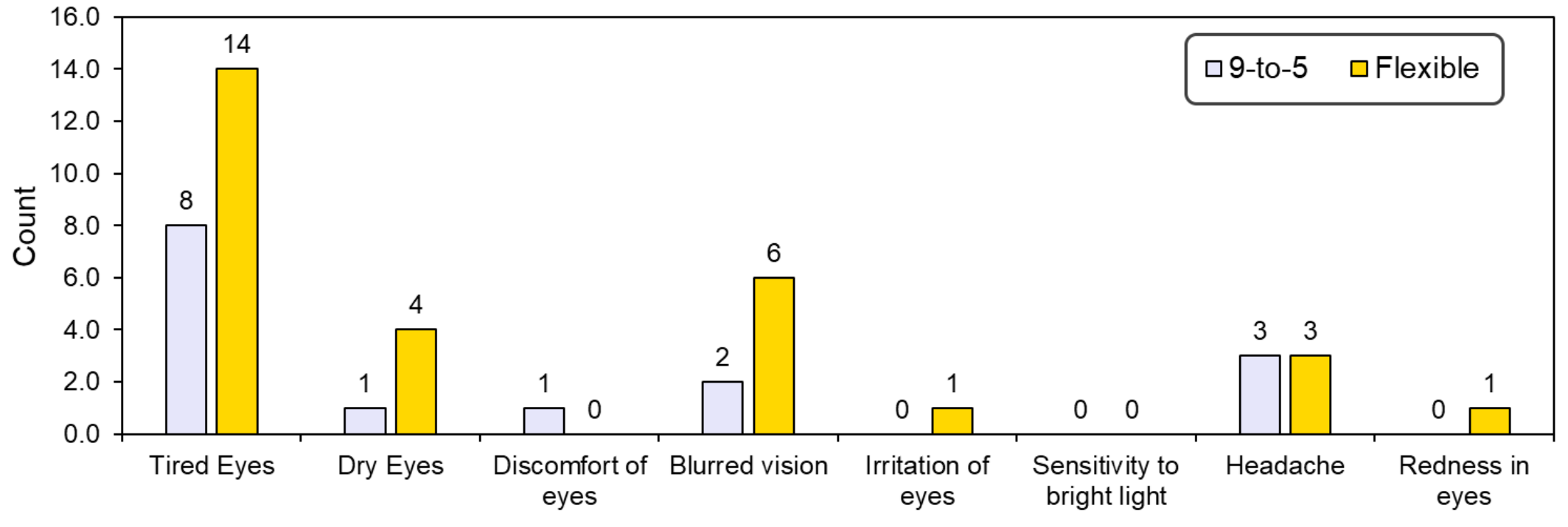}
\caption[The recorded eye-strain symptoms.]{The recorded eye-strain symptoms of participants after the test.}
\label{eye_strain_symptoms}
\end{figure}

Furthermore, we recorded participants' symptoms of eye-strain after each session as illustrated in Figure \ref{eye_strain_symptoms}. It should be noted that these symptoms are based on the subjective feedback from the participants and do not reflect clinical definitions. We found that the flexible group reported more eye-strain symptoms (29 symptoms) than the 9-to-5 group (15 symptoms). 

The large gap between groups is related to \textit{tired eyes}, \textit{dry eyes}, and \textit{blurred vision} which can be explained due to the higher working hours on a computer display of the flexible group as shown earlier in Figure \ref{display_hours_both}. In simple words, working in a `flexible' routine seems to lead to more hours spent on a display, which in turn leads to more eye-strain symptoms.

\begin{figure}[ht]
\centering
\includegraphics[width=0.75\textwidth]{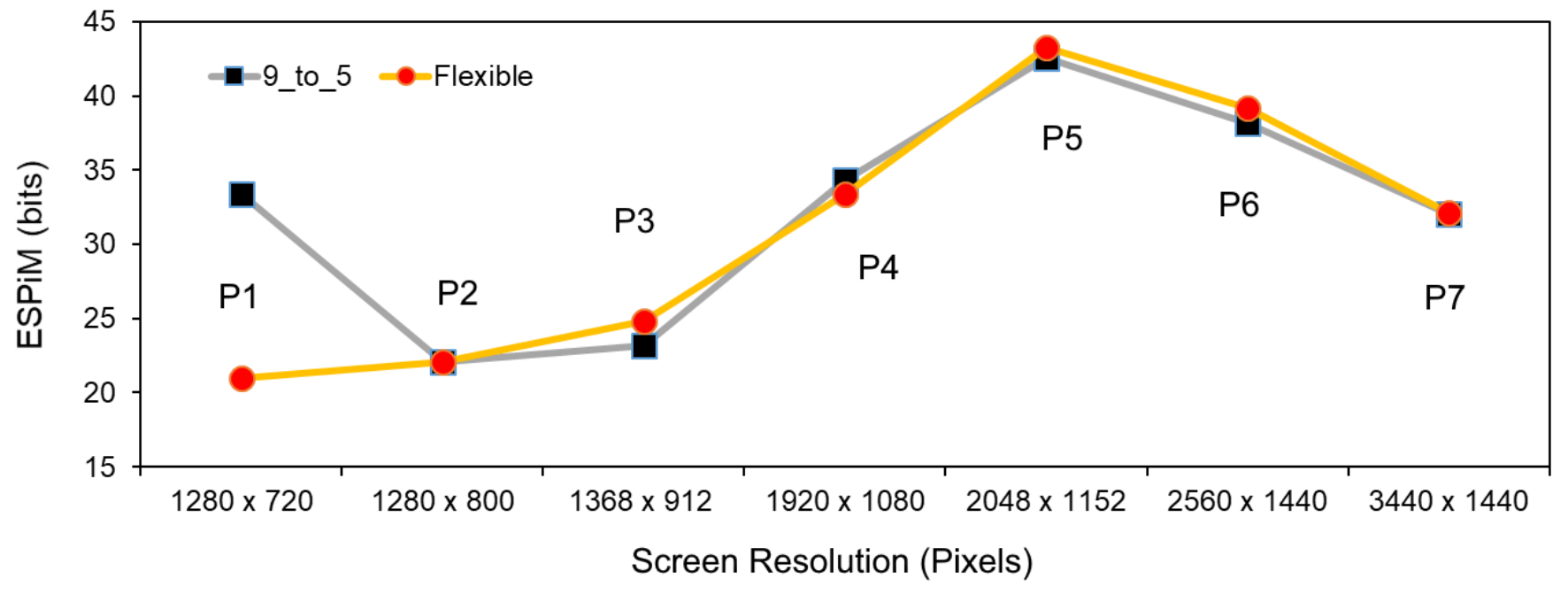}
\caption[ESPiM results based on screen resolutions.]{The recorded ESPiM values of both groups based on screen resolutions. The comparison points are shown as $P1$ to $P7$ which build comparison sections to study ESPiM on screen expansion. We analysed the increase and decrease of ESPiM in each section.}
\label{espim_screen_resolution}
\end{figure}

Since participants ran the test on their personal computers at home, we recorded different screen resolutions and analyzed the effectiveness of ESPiM on different screens as illustrated in Figure \ref{espim_screen_resolution}. We found no significant difference between the 9-to-5 group ($M=32.26~bits$,$SE=2.81~bits$) and the flexible group ($M=30.83~bits$,$SE=3.25~bits$) based on screen resolution, although there is a slight decrease of ESPiM for screen dimension $1280~\times~720$ pixels for the 9-to-5 group.  

Additionally, we calculated the difference of ESPiM values for each resolution point (shown as P1-P7 on Figure \ref{espim_screen_resolution}) to analyze the increase and decrease of ESPiM values per resolution increase. In fact, we calculated $P_{n}-P_{n-1}$ ($1 \leq n \land P_{0} = 0$) for each resolution as shown in Table \ref{tab:resolution_difference}. Negative values represent \textit{decrease}, and positive values \textit{increase} of ESPiM per bits. The summation of both groups shows a slightly higher increase for the flexible group with the value of $32.11~bits$ compared to that of the 9-to-5 group with the value of $32.01~bits$. 

However, the difference lies only on a single resolution ($1280~\times~720$~pixels), this might suggest that users with a flexible working time may experience higher eye-strain levels as screen dimensions increase than the group of standard 9-to-5 routine. 

However, the increase of eye-strain for both groups decreases for screen dimensions larger than $2048~\times~1152$ pixels. This may suggest that the choice of screen size is essential in how users experience eye-strain for any working time schedules. This result might be of interest for video game designers in finding suitable screen resolutions for their audience to recommend for the best game experience with lower eye-strain levels. 

\begin{table}[h!]
	\centering
	\caption[ESPiM difference based on screen resolution.]{ESPiM difference based on screen resolution for both 9-to-5, and flexible groups. Positive values represent increase and negative values decrease of ESPiM per bits respectively. }
	\label{tab:resolution_difference}
	{
		\begin{tabular}{|cccc|}
			\hline
			Resolution (Pixels) & Measure & ESPiM 9-to-5 & ESPiM Flexible \\
			\hline
			1280 $\times$ 720 	& P1 & 33.38 & 20.98 \\
			1280 $\times$ 800 	& P2 - P1 & -11.28 & 1.10 \\
			1368 $\times$ 912 	& P3 - P2 & 1.11  & 2.73 \\
			1920 $\times$ 1080 	& P4 - P3 & 11.13 & 8.58 \\
			2048 $\times$ 1152 	& P5 - P4 & 8.25 &  9.86 \\
			2560 $\times$ 1440 	& P6 - P5 & -4.41 & -4.03 \\
			3440 $\times$ 1440 	& P7 - P6 & -6.17 & -7.11 \\
			\hline
								& \textbf{$\Sigma$} & \textbf{32.01} & \textbf{32.11} \\
			\hline
		\end{tabular}
	}
\end{table}

This analysis is important as ESPiM takes both target's and screen's area into account to calculate eye-strain value for comparison as shown in Equation \ref{espim_equation}. Therefore, ESPiM is applicable on any screen resolution with no further adjustments in its equation. This feature enables designers of user interfaces, or producers of digital displays to compare different screen dimensions based on eye-strain on consumers.

\begin{figure}[ht]
\centering
\includegraphics[width=0.65\textwidth]{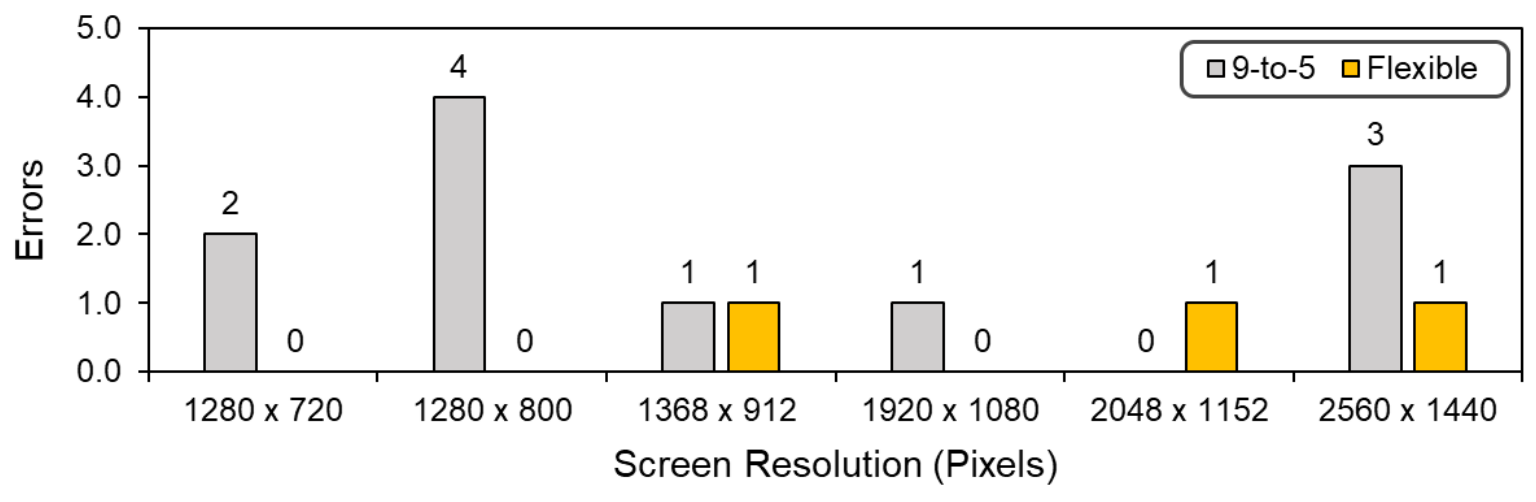}
\caption[Error rates based on screen resolutions.]{The recorded errors of both groups based on screen resolutions.}
\label{errors_screen_resolution}
\end{figure}

We also analyzed the error rates per screen resolution for both 9-to-5, and flexible groups as shown in Figure \ref{errors_screen_resolution}. We recorded 11 errors for the 9-to-5 group compared to 3 errors for the flexible group. This suggests a higher number of errors on smaller and larger screen dimensions for the 9-to-5 group.

We applied the \textit{K}-means clustering algorithm ($K=4$) of the \textit{Scikit-learn} library in Python \cite{scikit-learn} on the center of target locations on screen recorded for participants according to 4 equally-sized regions of the screen to analyze the relative distance of (a) fixation points, (b) mouse click coordinates, and (c) error locations recorded during the test. Since the mentioned measures were distributed throughout the entire screen (see Figure \ref{fixation_scatter} for eye fixation points) on different screen dimensions, we applied the clustering algorithm to downscale the locations into 4 points for comparison between measures of both 9-to-5, and flexible groups. 

\begin{figure}[ht]
\centering
\includegraphics[width=0.52\textwidth]{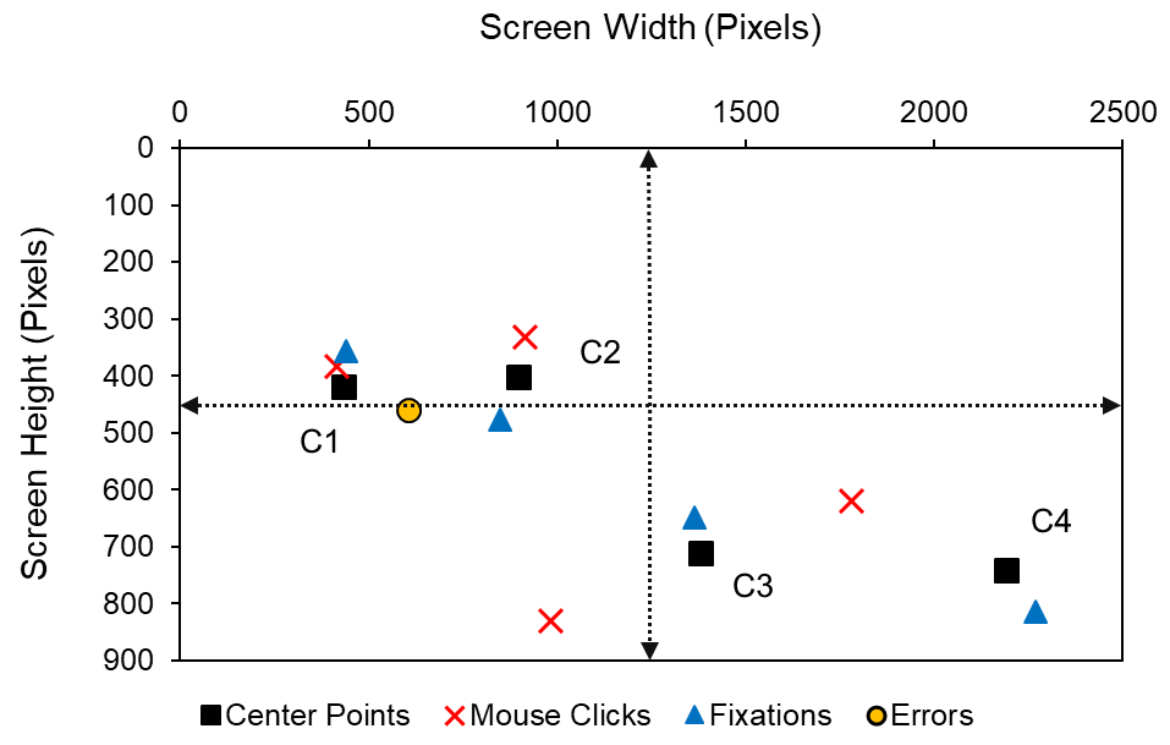}
\caption[Clusters 9-to-5]{The clusters of target center points ($C1$~to~$C4$) with the \textit{K}-means algorithm ($k=4$) for mouse clicks, fixation points and errors of the \textbf{9-to-5} group. The dashed lines represent imaginary division borders of screen into equal regions for comparison.}
\label{clusters_9_to_5}
\end{figure}

\begin{figure}[ht]
\centering
\includegraphics[width=0.52\textwidth]{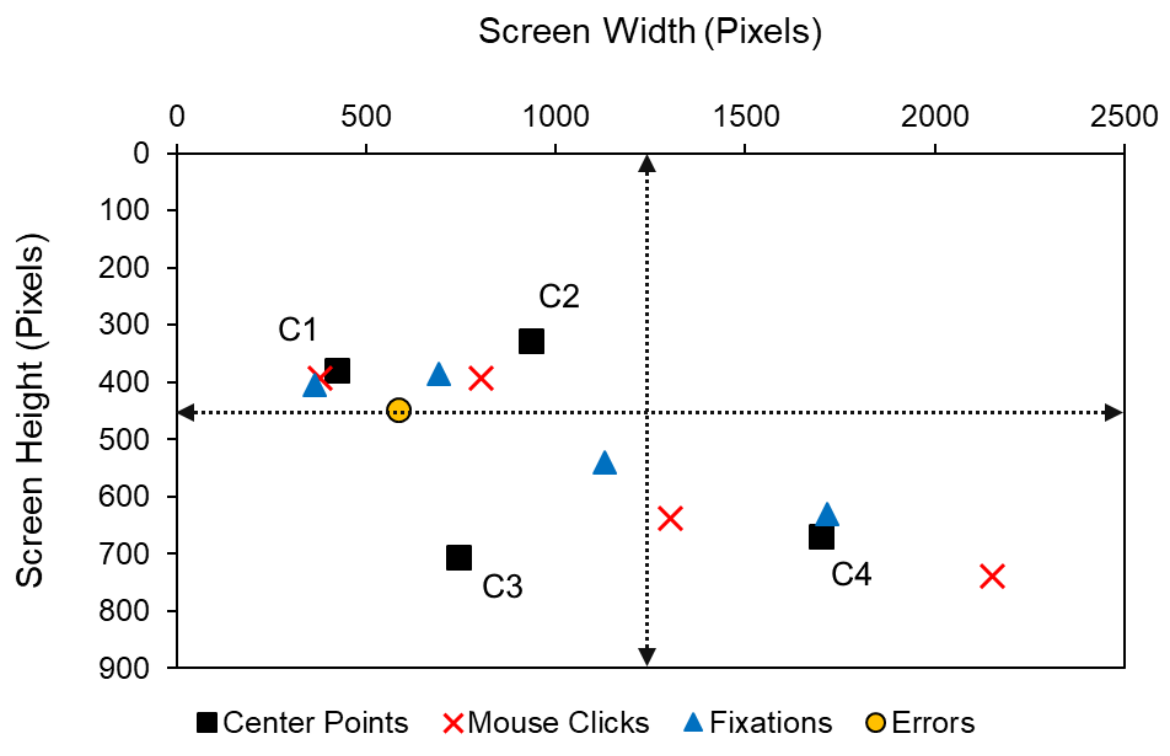}
\caption[Clusters flexible]{The clusters of target center points ($C1$~to~$C4$) with the \textit{K}-means algorithm ($k=4$) for mouse clicks, fixation points and errors of the \textbf{flexible} group. The dashed lines represent imaginary division borders of screen into equal regions for comparison.}
\label{clusters_flexible}
\end{figure}

The results are illustrated in Figure \ref{clusters_9_to_5}, \ref{clusters_flexible}. This analysis was applied of the center of targets illustrated as $C1$ to $C4$ on the plots to compare the proximity of each measure based on working times (9-to-5 vs. flexible). Interestingly, the locations of errors are almost similar for both groups. Although the flexible group made a lower error rate (3 errors) than the 9-to-5 group (11 errors), the clustering results show the same regions for both groups of participants. 

The upper left region shows a shorter distance for both fixations, and mouse clicks for both groups of participants labelled with $C1$, and $C2$ whereas the lower regions (left and right) show a higher drift from the target center points for mouse clicks illustrated as $C3$ and $C4$ on Figure \ref{clusters_9_to_5}, \ref{clusters_flexible}. In addition, the lower regions of the 9-to-5 group show a short distance to target centers for eye fixations (Figure \ref{clusters_9_to_5}), whereas the left lower region shows a larger drift from the target center $C3$ for the flexible group as shown in Figure \ref{clusters_flexible}.
 
\begin{figure}[ht]
\centering
 \subfigure{\includegraphics[width=0.2\textwidth]{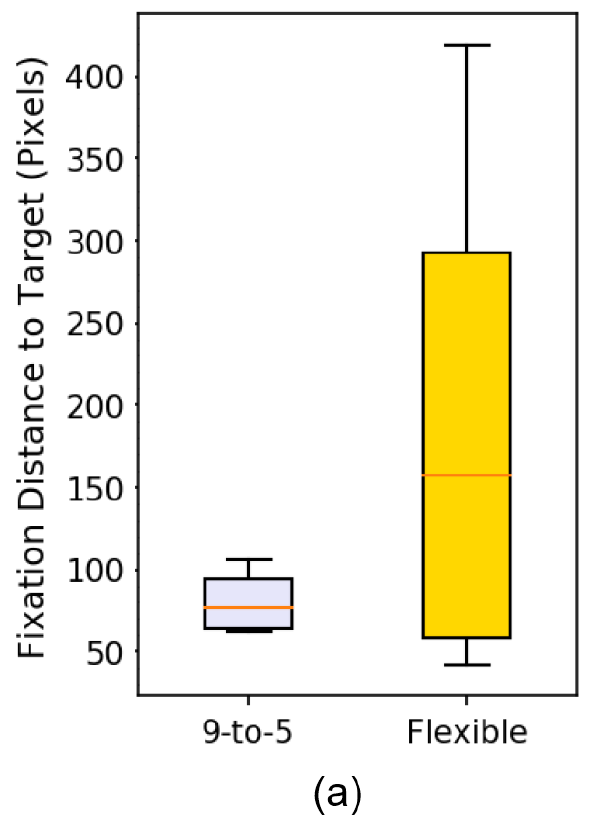}
 \label{fixation_distance}}
 \subfigure{\includegraphics[width=0.2\textwidth]{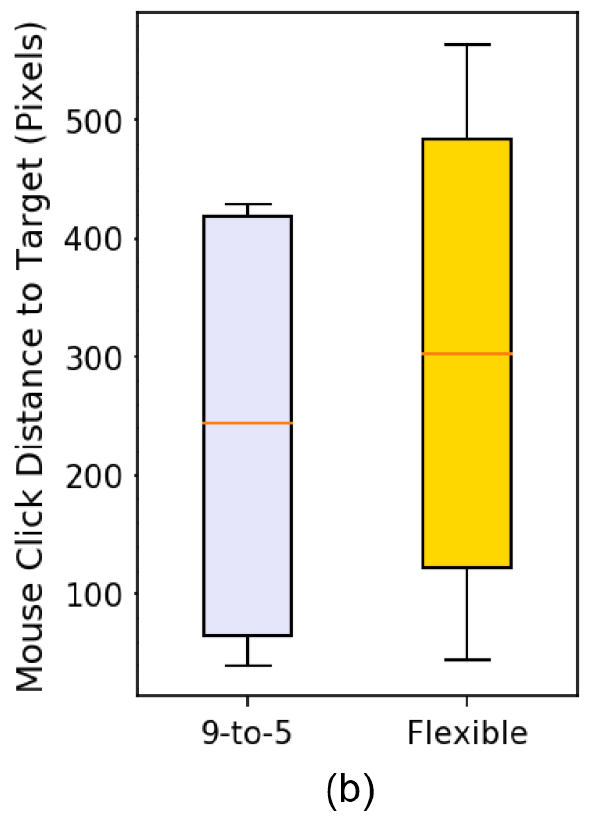}
 \label{mouse_clicks_distance}}
\caption[Euclidean distance plots.]{The recorded Euclidean distance to target centers ($p > .05$) for (a) fixation points, and (b) mouse clicks.}
\label{both_group_distances}
\end{figure}

Furthermore, we calculated the Euclidean distance to target centers for (a) fixation points, and (b) mouse click coordinates for both groups to compare the distance drift of each group from the center of targets as illustrated in Figure \ref{both_group_distances}. This suggests, that the flexible group recorded a higher distance to target centers than the 9-to-5 group, however, the difference is statistically not significant ($p > .05$), which may lead to a lower selection accuracy based on center points.

We demonstrated the application of ESPiM in a remote user study with a video-based eye-tracking technique. We have shown the effectiveness of ESPiM for user studies to analyze eye-strain on digital displays. Although the results of ESPiM cannot be interpreted as clinical analysis, they can be applied in low-budged user studies with physical restrictions to participants.

\section{Limitations and Potentials of ESPiM}
Although the ESPiM model is designed to be as flexible as possible, a precise measurement of test execution time is required for any user study. Thus an accurate synchronization between the start and end of a test is needed. Further, it does not take preparation activity times into account, for instance, spent time on the calibration process before running a test.
Moreover, the ESPiM model represents a theoretical model for eye-strain, however, this model does not represent medical/clinical measures and therefore is not applicable in these fields of study.

Beyond the applications of ESPiM mentioned before in section \ref{applications-of-ESPiM}, this model can also be suitable for analyzing Microsleep, a sudden loss of awareness or a short sleep in car drivers. Especially for truck drivers, this could reduce traffic collisions by measuring eye-strain levels for different times and conditions in user studies. Researchers may apply machine learning techniques to run or expand the model in real-time use case scenarios to detect driver's awareness based on eye-strain through integrated eye-tracking systems into car dashboards. In addition, the ESPiM model can be applied in research studies involving wearable eye-tracking devices, and biological sensors to study the impacts of visual distractions among knowledge workers.

\section{Conclusion and Future Work}

Eye-strain is a common issue among computer users due to prolonged periods spent working and using digital displays, which leads to vision problems such as irritation and tiredness of the eyes and headaches. We proposed the Eye-Strain Probation Model (ESPiM), an easy-to-apply computational model to measure eye-strain on digital displays based on the spatial properties of the user interface and display area for a required period of time, using eye-tracking analysis integrated into a single measure. 

We conducted two user studies to evaluate the effectiveness of ESPiM, its functionalities and potentials and showed how to measure potential eye-strain levels of a specific user interface suitable for pilot studies to compare various design prototypes before the application by end-users.

We evaluated the effectiveness of ESPiM in an in-person user study with an infrared eye-tracking sensor and found interesting patterns among video gameplay play frequency groups. In addition, we showed the application of ESPiM in a remote user study that complied with the COVID-19 safety measures. Since eye fixations can be assessed, which is typically bound to the range of (200-600~ms), ESPiM can be applied in pilot studies without access to eye-tracking devices for analysis estimations. 

Despite the relatively short test session for each participant of both user studies, we evaluated our hypotheses and identified distinctive patterns among participants. Furthermore, we showed that ESPiM has strong correlations with the \textit{error rates} of target selections. The correlation with error rates can be interpreted as an indicator to estimate and reduce the impacts of the Midas touch problem in gaze-based interactions by analyzing the user interface properties. 

We also found significantly different eye-strain patterns based on the video gameplay frequency of participants. The results showed that participants with frequent video gameplay reached significantly lower eye-strain and lower error rates than their counterparts. This may suggest that users with higher training of eye focus shifts (e.g. video games) might experience a lower eye-strain with prolonged use of digital displays for singular (not multi-tasking) or enjoyable tasks. Furthermore, we found that mouse pointer movements increase as eye-strain levels increase. This would suggest users tend to move their mouse pointers more frequently to select targets in case of tired eyes. Beyond the prediction and measurement of eye-strain, ESPiM can be applied to compare different gaze-based interaction techniques and evaluate different user interface prototypes to reach a comfortable design based on eye-strain. 

As more individuals get access to digital content provided on digital displays, especially children and students of a younger age, digital media consumption becomes more prevalent. Furthermore, since the start of the COVID-19 pandemic, many schools have moved to online teaching, which has caused new challenges for elementary students and their parents. Many in-person events still occur on digital devices now to comply with the required safety measures. Thus, there is a need for compound models to cope with the emerging trends in order to measure and compare the impact of digital displays on our health. Today, most smartphones include high-quality cameras accessible by younger generations. Thus the application of video-based eye-tracking that we showed in this paper can be feasible to design and conduct large-scale user studies on smartphones to reach a larger population of digital media consumers. 

In future work, we plan to apply ESPiM on smart-phones to compare educational video games for school students. Additionally, the continuous and predictable form of ESPiM makes it suitable for machine learning algorithms which will be explored in the future. Finally, we hope our proposed model makes a step forward towards reducing eye-strain on digital displays, inspires researchers and user interface designers and leads to more discussions in research communities.


\bibliographystyle{plain}
\bibliography{references}

\begin{thebibliography}{10}

\bibitem{abdulin2015user}
Evgeniy Abdulin and Oleg Komogortsev.
\newblock User eye fatigue detection via eye movement behavior.
\newblock In {\em Proceedings of the 33rd annual ACM conference extended
  abstracts on human factors in computing systems}, pages 1265--1270. ACM,
  2015.

\bibitem{achtman2008video}
Rebecca~L Achtman, C~Shawn Green, and Daphne Bavelier.
\newblock Video games as a tool to train visual skills.
\newblock {\em Restorative neurology and neuroscience}, 26(4, 5):435--446,
  2008.

\bibitem{aws_Elastic}
Inc. or its~affiliates Amazon Web~Services.
\newblock Aws elastic beanstalk.
\newblock \url{https://aws.amazon.com/elasticbeanstalk/}, 2021.

\bibitem{bahill1975overlapping}
A~Terry Bahill and Lawrence Stark.
\newblock Overlapping saccades and glissades are produced by fatigue in the
  saccadic eye movement system.
\newblock {\em Experimental neurology}, 48(1):95--106, 1975.

\bibitem{card1978evaluation}
Stuart~K Card, William~K English, and Betty~J Burr.
\newblock Evaluation of mouse, rate-controlled isometric joystick, step keys,
  and text keys for text selection on a crt.
\newblock {\em Ergonomics}, 21(8):601--613, 1978.

\bibitem{CASTEL2005217}
Alan~D. Castel, Jay Pratt, and Emily Drummond.
\newblock The effects of action video game experience on the time course of
  inhibition of return and the efficiency of visual search.
\newblock {\em Acta Psychologica}, 119(2):217 -- 230, 2005.

\bibitem{covid_19_intro}
Marco Ciotti, Massimo Ciccozzi, Alessandro Terrinoni, Wen-Can Jiang, Cheng-Bin
  Wang, and Sergio Bernardini.
\newblock The covid-19 pandemic.
\newblock {\em Critical reviews in clinical laboratory sciences},
  57(6):365--388, 2020.

\bibitem{crossman1983feedback}
ERFW Crossman and PJ~Goodeve.
\newblock Feedback control of hand-movement and fitts' law.
\newblock {\em The Quarterly Journal of Experimental Psychology Section A},
  35(2):251--278, 1983.

\bibitem{di2013saccadic}
Leandro~L Di~Stasi, M~Marchitto, A~Antol{\'\i}, and Jos{\'e}~J Ca{\~n}as.
\newblock Saccadic peak velocity as an alternative index of operator attention:
  A short review.
\newblock {\em Revue Europ{\'e}enne de Psychologie Appliqu{\'e}e/European
  Review of Applied Psychology}, 63(6):335--343, 2013.

\bibitem{fitts1954information}
Paul~M Fitts.
\newblock The information capacity of the human motor system in controlling the
  amplitude of movement.
\newblock {\em Journal of experimental psychology}, 47(6):381, 1954.

\bibitem{gori2017one}
Julien Gori, Olivier Rioul, Yves Guiard, and Michel Beaudouin-Lafon.
\newblock One fitts’ law, two metrics.
\newblock In {\em IFIP Conference on Human-Computer Interaction}, pages
  525--533. Springer, 2017.

\bibitem{hansen2018fitts}
John~Paulin Hansen, Vijay Rajanna, I~Scott MacKenzie, and Per B{\ae}kgaard.
\newblock A fitts' law study of click and dwell interaction by gaze, head and
  mouse with a head-mounted display.
\newblock In {\em Proceedings of the Workshop on Communication by Gaze
  Interaction}, pages 1--5, 2018.

\bibitem{isomoto2018dwell}
Toshiya Isomoto, Toshiyuki Ando, Buntarou Shizuki, and Shin Takahashi.
\newblock Dwell time reduction technique using fitts' law for gaze-based target
  acquisition.
\newblock In {\em Proceedings of the 2018 ACM Symposium on Eye Tracking
  Research \& Applications}, pages 1--7, 2018.

\bibitem{iwa2005image}
ISO IWA.
\newblock {ISO}, image safety: Reducing the incidence of undesirable biomedical
  effects caused by visual image sequence, 2005.

\bibitem{JASP2019}
{JASP Team}.
\newblock {JASP (Version 0.11.1)[Computer software]}, 2019.

\bibitem{keele1968processing}
Steven~W Keele and Michael~I Posner.
\newblock Processing of visual feedback in rapid movements.
\newblock {\em Journal of experimental psychology}, 77(1):155, 1968.

\bibitem{kim2014effect}
Heejin Kim, Sunghyuk Kwon, Jiyoon Heo, Hojin Lee, and Min~K Chung.
\newblock The effect of touch-key size on the usability of in-vehicle
  information systems and driving safety during simulated driving.
\newblock {\em Applied ergonomics}, 45(3):379--388, 2014.

\bibitem{komogortsev2010standardization}
Oleg~V Komogortsev, Denise~V Gobert, Sampath Jayarathna, Do~Hyong Koh, and
  Sandeep~M Gowda.
\newblock Standardization of automated analyses of oculomotor fixation and
  saccadic behaviors.
\newblock {\em IEEE Transactions on Biomedical Engineering}, 57(11):2635--2645,
  2010.

\bibitem{kuze2008subjective}
J~Kuze and K~Ukai.
\newblock Subjective evaluation of visual fatigue caused by motion images.
\newblock {\em Displays}, 29(2):159--166, 2008.

\bibitem{lanthier2013measuring}
Sophie Lanthier, Evan Risko, Daniel Smilek, and Alan Kingstone.
\newblock Measuring the separate effects of practice and fatigue on eye
  movements during visual search.
\newblock In {\em Proceedings of the Annual Meeting of the Cognitive Science
  Society}, volume~35, 2013.

\bibitem{screen_time_risks}
Gadi Lissak.
\newblock Adverse physiological and psychological effects of screen time on
  children and adolescents: Literature review and case study.
\newblock {\em Environmental Research}, 164:149--157, 2018.

\bibitem{MACK201426}
David~J. Mack and Uwe~J. Ilg.
\newblock The effects of video game play on the characteristics of saccadic eye
  movements.
\newblock {\em Vision Research}, 102:26 -- 32, 2014.

\bibitem{mackenzie1989note}
I~Scott MacKenzie.
\newblock A note on the information-theoretic basis for fitts’ law.
\newblock {\em Journal of motor behavior}, 21(3):323--330, 1989.

\bibitem{mackenzie2012evaluating}
I~Scott MacKenzie.
\newblock Evaluating eye tracking systems for computer input.
\newblock In {\em Gaze interaction and applications of eye tracking: Advances
  in assistive technologies}, pages 205--225. IGI Global, 2012.

\bibitem{mackenzie1992extending}
I~Scott MacKenzie and William Buxton.
\newblock Extending fitts' law to two-dimensional tasks.
\newblock In {\em Proceedings of the SIGCHI conference on Human factors in
  computing systems}, pages 219--226, 1992.

\bibitem{majaranta2009fast}
P{\"a}ivi Majaranta, Ulla-Kaija Ahola, and Oleg {\v{S}}pakov.
\newblock Fast gaze typing with an adjustable dwell time.
\newblock In {\em Proceedings of the SIGCHI Conference on Human Factors in
  Computing Systems}, pages 357--360. ACM, 2009.

\bibitem{majaranta2014eye}
P{\"a}ivi Majaranta and Andreas Bulling.
\newblock Eye tracking and eye-based human--computer interaction.
\newblock In {\em Advances in physiological computing}, pages 39--65. Springer,
  2014.

\bibitem{megaw1983visual}
Ted Megaw and Tayyar Sen.
\newblock Visual fatigue and saccadic eye movement parameters.
\newblock In {\em Proceedings of the Human Factors Society Annual Meeting},
  volume~27, pages 728--732. Sage Publications Sage CA: Los Angeles, CA, 1983.

\bibitem{screen_time_children}
Yui Mineshita, Hyeon-Ki Kim, Hanako Chijiki, Takuya Nanba, Takae Shinto, Shota
  Furuhashi, Satoshi Oneda, Mai Kuwahara, Anzu Suwama, and Shigenobu Shibata.
\newblock Screen time duration and timing: effects on obesity, physical
  activity, dry eyes, and learning ability in elementary school children.
\newblock {\em BMC public health}, 21(1):1--11, 2021.

\bibitem{IDEA}
{Mohsen Parisay}, Charalambos Poullis, and Marta Kersten-Oertel.
\newblock Idea: Index of difficulty for eye tracking applications - an analysis
  model for target selection tasks.
\newblock In {\em Proceedings of the 16th International Joint Conference on
  Computer Vision, Imaging and Computer Graphics Theory and Applications -
  Volume 1: HUCAPP}, pages 135--144. INSTICC, SciTePress, 2021.

\bibitem{mowatt2018computer}
Lizette Mowatt, Carron Gordon, Arvind Babu~Rajendra Santosh, and Thaon Jones.
\newblock Computer vision syndrome and ergonomic practices among undergraduate
  university students.
\newblock {\em International journal of clinical practice}, 72(1):e13035, 2018.

\bibitem{munshi2017computer}
Sunil Munshi, Ashley Varghese, and Sushma Dhar-Munshi.
\newblock Computer vision syndrome—a common cause of unexplained visual
  symptoms in the modern era.
\newblock {\em International Journal of Clinical Practice}, 71(7):e12962, 2017.

\bibitem{papoutsaki2016webgazer}
Alexandra Papoutsaki, Patsorn Sangkloy, James Laskey, Nediyana Daskalova, Jeff
  Huang, and James Hays.
\newblock Webgazer: Scalable webcam eye tracking using user interactions.
\newblock In {\em Proceedings of the 25th International Joint Conference on
  Artificial Intelligence (IJCAI)}, pages 3839--3845. AAAI, 2016.

\bibitem{felix}
Mohsen Parisay, Charalambos Poullis, and Marta Kersten-Oertel.
\newblock Felix: Fixation-based eye fatigue load index a multi-factor measure
  for gaze-based interactions.
\newblock In {\em 2020 13th International Conference on Human System
  Interaction (HSI)}, pages 74--81, 2020.

\bibitem{EyeTAP}
Mohsen Parisay, Charalambos Poullis, and Marta Kersten-Oertel.
\newblock Eyetap: Introducing a multimodal gaze-based technique using voice
  inputs with a comparative analysis of selection techniques.
\newblock {\em International Journal of Human-Computer Studies}, 154:102676,
  2021.

\bibitem{scikit-learn}
F.~Pedregosa, G.~Varoquaux, A.~Gramfort, V.~Michel, B.~Thirion, O.~Grisel,
  M.~Blondel, P.~Prettenhofer, R.~Weiss, V.~Dubourg, J.~Vanderplas, A.~Passos,
  D.~Cournapeau, M.~Brucher, M.~Perrot, and E.~Duchesnay.
\newblock Scikit-learn: Machine learning in {P}ython.
\newblock {\em Journal of Machine Learning Research}, 12:2825--2830, 2011.

\bibitem{rosenfield2016computer}
Mark Rosenfield.
\newblock Computer vision syndrome (aka digital eye strain).
\newblock {\em Optometry in Practice}, 17(1):1--10, 2016.

\bibitem{vspakov2004line}
Oleg {\v{S}}pakov and Darius Miniotas.
\newblock On-line adjustment of dwell time for target selection by gaze.
\newblock In {\em Proceedings of the third Nordic conference on Human-computer
  interaction}, pages 203--206. ACM, 2004.

\bibitem{dry_eyes_definition}
Kazuo Tsubota.
\newblock Tear dynamics and dry eye.
\newblock {\em Progress in Retinal and Eye Research}, 17(4):565--596, 1998.

\bibitem{ukai2008visual}
Kazuhiko Ukai and Peter~A Howarth.
\newblock Visual fatigue caused by viewing stereoscopic motion images:
  Background, theories, and observations.
\newblock {\em Displays}, 29(2):106--116, 2008.

\bibitem{vasiljevas2016modelling}
Mindaugas Vasiljevas, T~Gedminas, A~{\v{S}}ev{\v{c}}enko, M~Jan{\v{c}}iukas,
  T~Bla{\v{z}}auskas, and R~Dama{\v{s}}evi{\v{c}}ius.
\newblock Modelling eye fatigue in gaze spelling task.
\newblock In {\em 2016 IEEE 12th International Conference on Intelligent
  Computer Communication and Processing (ICCP)}, pages 95--102. IEEE, 2016.

\bibitem{wobbrock2011effects}
Jacob~O Wobbrock, Kristen Shinohara, and Alex Jansen.
\newblock The effects of task dimensionality, endpoint deviation, throughput
  calculation, and experiment design on pointing measures and models.
\newblock In {\em Proceedings of the SIGCHI Conference on Human Factors in
  Computing Systems}, pages 1639--1648. ACM, 2011.

\end{thebibliography}

\end{document}